# Differential velocity cumulative distribution (DVCD): an efficient statistical method for binary-star analysis in large stellar catalogues using radial velocities

Feng Luo ,[1,2] Yong-Heng Zhao[1,2★] and Chao Liu [1,2]

[1] *National Astronomical Observatories, Chinese Academy of Sciences, Beijing 100012, China*
[2] *University of Chinese Academy of Sciences, Beijing 100049, China*

Accepted 2026 March 20. Received 2026 March 18; in original form 2025 November 22

## ABSTRACT

Binary stars are fundamental to astrophysics, offering crucial insights into stellar evolution, galactic dynamics, and fundamental physics. Nevertheless, the high dimensionality of orbital parameters and observational constraints poses significant challenges for statistically characterizing their properties. In this study, we present a novel algorithm called the differential velocity cumulative distribution (DVCD) for analysing binary-star systems using radial velocity data. The DVCD method exhibits superior accuracy and computational efficiency compared to existing approaches, achieving computation time reductions of $10^{-4}$ to $10^{-5}$ under equivalent conditions. We applied the DVCD algorithm to red giant samples from APOGEE DR16, dividing the data set into 16 subsets based on $\log g$ and M/H. Our findings reveal that the intrinsic binary fraction decreases with decreasing surface gravity and increasing metallicity, offering valuable constraints on the evolutionary processes of binary stars. This study underscores the potential of the DVCD method for large-scale statistical analyses of binary systems.

## 1 INTRODUCTION

Binary stars represent a pivotal component of astronomical and astrophysical research. The gravitational and radiative interactions between components of close binary systems provide stringent diagnostics of fundamental stellar properties and evolutionary pathways (G. Peters 2011; H. Ge et al. 2015; R. S. Khoshnood et al. 2015; T. Kangas et al. 2017; J. R. Lomax et al. 2017; D. R. Gies et al. 2020; M. A. M. Kooten, M. Kenworthy & N. Doelman 2020; C. N. Mazzola et al. 2020). Binary systems are associated with a diverse array of astrophysical phenomena, including X-ray binaries (F. García et al. 2022; S. Karino 2022; S. Wang et al. 2022; W.-L. Zhang et al. 2022), supernovae (D. Tsebrenko & N. Soker 2015; P. M. Saz Parkinson et al. 2016; E. Ziegerer et al. 2017; Y. Shao & X.-D. Li 2018; S. Blondin et al. 2022; J. Shin et al. 2021), and kilonovae (T. A. Callister 2021; R. K. Dubey & S. D. Pathak 2021). These systems function as natural laboratories for testing fundamental physical theories under extreme conditions, such as elevated temperatures, intense magnetic fields, and strong gravitational environments. For example, the gravitational wave event GW150914, resulting from the merger of a black hole binary system, provided compelling evidence for general relativity and validated Hawking's black hole area theorem (M. Isi et al. 2021).

The study of binary stars is also crucial for understanding stellar evolution. Binary interactions, such as mass transfer, common-envelope evolution, and mergers, significantly influence the life cycles of stars, leading to phenomena such as blue stragglers, Type Ia supernovae, and compact-object binaries (H. Sana et al. 2012; A. Sota et al. 2014; C. N. Mazzola et al. 2020). These processes play a key role in shaping the stellar populations of galaxies and contribute to the chemical enrichment of the interstellar medium. In particular, red giant binaries provide a unique opportunity to study late-stage stellar evolution, as their extended envelopes and slow rotation rates make them sensitive probes of binary interactions (C. Badenes et al. 2018).

Radial velocity (RV) measurements, derived from the Doppler effect, constitute a powerful diagnostic tool for investigating the kinematics of celestial bodies. In binary systems, periodic variations in RVs encode information about orbital parameters such as periods, masses, eccentricities, and inclinations. While sufficient observational data enable precise determination of these parameters, limited data often necessitate statistical approaches. Statistical analyses of binaries provide insights into the evolution of stars, binary systems, and galaxies. Recent studies have focused on binary fractions (R. Chini et al. 2012; H. Sana et al. 2012; H. A. Kobulnicky et al. 2014; A. Sota et al. 2014; E. J. Aldoretta et al. 2015; P. R. Dunstall et al. 2015), but the distribution of orbital parameters remains less well understood (H. Sana et al. 2013; S. Gao et al. 2014, 2017; G. Yanjun et al. 2022; F. Luo et al. 2021). Challenges include limited observational data, selection biases, and methodological constraints that affect the accuracy of derived results.

★ E-mail: yzhao@bao.ac.cn

**Table 1.** Orbital parameters that govern the radial velocity variations of binary stars.

| Name | Description |
|---|---|
| $P$ | Orbital period |
| $q$ | Mass ratio |
| $e$ | Eccentricity |
| $m_1$ | Mass of the primary |
| $i$ | Inclination angle |
| $\omega$ | Argument of periastron |
| $T_0$ | Time of periastron passage |

The APOGEE survey, part of the Sloan Digital Sky Survey, has provided high-resolution, near-infrared spectra for hundreds of thousands of stars, including red giants, across the Milky Way (S. R. Majewski et al. 2017; A. M. Price-Whelan et al. 2020). This data set offers an unprecedented opportunity to study binary populations in different Galactic environments. However, the analysis of binary stars in APOGEE data is complicated by the limited number of RV measurements per star, the presence of observational noise, and the need to disentangle binary-induced RV variations from intrinsic stellar variability (N. W. Troup et al. 2016).

In this investigation, we introduce the differential velocity cumulative distribution (DVCD) algorithm, a novel statistical method designed to estimate intrinsic binary fractions ($f_{\rm bin}$, hereafter) and predict orbital parameter distributions from RV data. Unlike traditional methods, DVCD leverages the cumulative distribution characteristics of RV difference sequences, enabling robust statistical inference even in the presence of sparse and noisy data. The algorithm is computationally efficient, achieving speedups of several orders of magnitude compared to existing approaches, making it well-suited for large-scale surveys like APOGEE.

Empirical results demonstrate the algorithm's exceptional accuracy and computational efficiency. We apply DVCD to red-giant samples from APOGEE DR16, partitioned into 16 bins based on surface gravity (logg) and metallicity ([M/H]). Our analysis reveals that the $f_{\rm bin}$ decreases with decreasing surface gravity and increasing metallicity, providing new constraints on binary interaction and evolutionary processes in red giants. These findings underscore the potential of DVCD for advancing our understanding of binary-star populations and their role in stellar and galactic evolution.

This manuscript is organized as follows: Section 1 provides an introduction to the research context. Section 2 describes the binarity criterion. Section 3 discusses the cumulative distribution characteristics of radial velocity difference sequences. Section 4 elaborates on the DVCD algorithm implementation. Section 5 presents the application to APOGEE DR16 data. Finally, Section 6 discusses the results and conclusions.

## 2 CRITERION FOR BINARITY

The RV variations of binary stars are governed by seven orbital parameters, as summarized in Table 1. The RV equation (equation 1) is derived from Kepler's laws and elliptical geometry (R. W. Hilditch 2001).

$$\mathrm{RV} = \frac{2\pi a_1 \sin i}{P\sqrt{(1-e^2)}}[\cos(\theta+\omega)+e\cos\omega] + \gamma,$$
$$a_1 = \frac{a}{(1+q)},\ a = \left(\frac{Gm_1(1+q)P^2}{4\pi^2}\right)^{\frac{1}{3}},$$
$$\theta = \cos^{-1}\left(\frac{\cos E - e}{1-e\cos E}\right),\ E - e\sin E = 2\pi\frac{t-T_0}{P}. \quad (1)$$

Here, $\gamma$ represents the systemic velocity, $a_1$ is the long axis of the elliptical orbit of the primary, which is associated with the mass ratio $q$, and $\theta$ is the true anomaly. Simulated RVs can be generated by specifying these orbital parameters and observation epochs.

To extract the orbital parameters from the observed RV data, the first step is to remove the systemic velocity $\gamma$. When the number of observations is sufficiently large – such that the orbital phase sampling spans at least one full orbital cycle – the systemic velocity can be reliably estimated as the arithmetic mean of all RV measurements. However, in practical observational scenarios, the orbital period is typically an unknown parameter and must itself be inferred from the data. To address this limitation, we adopt the sequence of chronological radial velocity differences (hereafter referred to as the difference sequence, DS) as the primary observable. Specifically, the DS comprises the differences between consecutive RV measurements, ordered chronologically by observation time.

Each element in the DS is denoted as ΔAJRV (adjacent radial velocity). For a source with $n$ repeated RV measurements, these values are first sorted chronologically by observation time. Then, each measurement – except the last one – is subtracted from the subsequent adjacent RV measurement, resulting in a ΔAJRV sequence containing $n-1$ elements:

$$\Delta\mathrm{AJRV}_j = \mathrm{RV}_k - \mathrm{RV}_j, \quad (j = 1, 2, 3, \ldots, n-1;\ k = j+1), \quad (2)$$

so that each $\Delta\mathrm{AJRV}_j$ is invariant to an additive systemic offset and encodes the relative temporal change of the radial velocity. This approach effectively mitigates potential systematic errors arising from uncertainties in systemic velocity determination.

Due to the influence of factors such as instrument characteristics, spectral resolution, SNR, and RV measurement methods, there will always be certain errors in the RV measurement of a particular observed spectrum. Assume that the measurement error of RV follows a normal distribution with a mean of 0 and a standard deviation of $\varepsilon$, RV $\sim N(0, \varepsilon)$. It can be known from probability theory that ΔAJRV also follows a normal distribution with a mean of 0 and a standard deviation of $\sqrt{2}\varepsilon$, that is ΔAJRV $\sim N(0, \sqrt{2}\varepsilon)$. Accordingly, we construct the likelihood function as presented in equation (3) to estimate the orbital parameters of a given binary system:

$$L(\theta^*|\Delta\mathrm{AJRV}) = \prod f(\Delta\mathrm{AJRV}_j|\theta^*),$$
$$f(\Delta\mathrm{AJRV}_j|\theta^*) = N(\Delta\mathrm{AJRV}_{j_{\rm simul}}, \mu = \Delta\mathrm{AJRV}_{j_{\rm obs}}, \sigma = \sqrt{2}\varepsilon),$$
$$\theta^* = (P, q, e, m_1, i, \omega, T_0). \quad (3)$$

The term of ΔAJRV subscripted simul is the simulated data generated by RV equation (1), while the term subscripted obs is derived from the observational RVs.

We generated a mock binary orbit sample with $\gamma = 50$, $\varepsilon = 1.0$, $\theta^* = (300, 0.7, 0.4, 6, \pi/3, \pi/6, 90)$ and randomly sampled the RV twenty times. The MCMC sampling was performed using the emcee Python package, employing 10 000 walkers and 10 000 steps per walker to ensure robust convergence. The burn-in phase consisted of the first 30 000 steps, which were discarded to mitigate initialization effects. The detection ranges of parameters for this experiment are specified in Table 2. These ranges were chosen based on prior knowledge of binary-star orbital parameters, ensuring that the true values of $P$, $q$, $e$, and $m_1$ were well within the explored parameter space. For $i$, and $\omega$, the ranges were set to cover all possible orientations and also avoid the occurrence of multiple extreme values because of the periodicity of trigonometric functions. The time of periastron passage $T_0$ was constrained to be within one orbital period to reflect the periodic nature of the system.

**Table 2.** Detection ranges of orbital parameters.

| Name | Range |
|---|---|
| $P$ (d) | [1, 600] |
| $q$ | [0.1, 1] |
| $e$ | [0, 0.9] |
| $m_1$ ($M_\odot$) | [1, 10] |
| $i$ | [0, $+\pi$] |
| $\omega$ | [$-\pi$, $+\pi$] |
| $T_0$ | [0, $P$] |

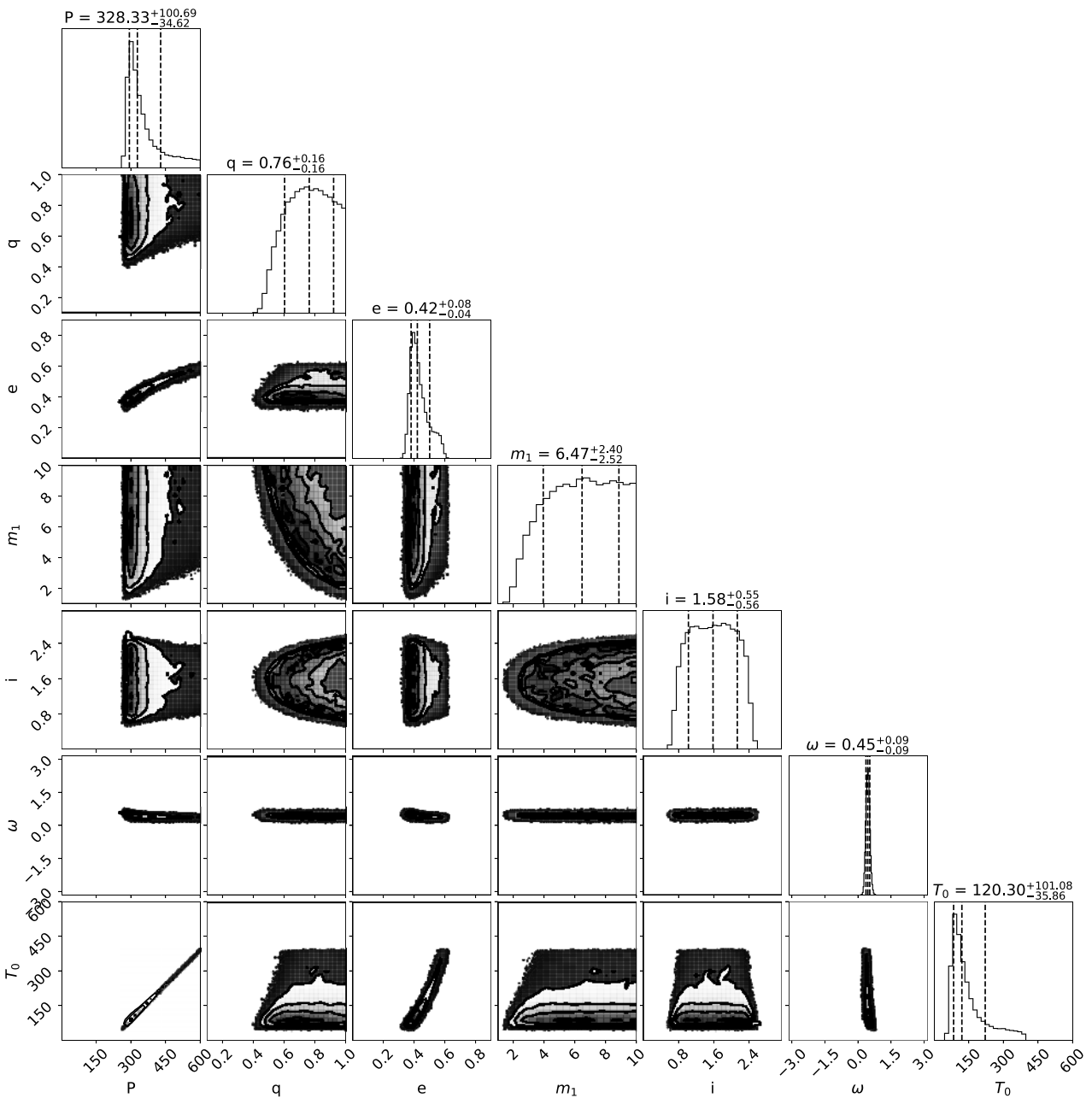

**Figure 1.** Result of the prediction of orbital parameters using the MCMC approach. The black dashed lines in each 1D subgraph respectively indicate the position of the 15th, 50th, and 85th percentiles.

Results obtained using the emcee Python package are illustrated in Fig. 1. For enhanced clarity, data points with likelihood probabilities below the 10th percentile have been excluded from the visualization. Analysis reveals that the eccentricity $e$ and argument of periastron $\omega$ were well constrained within narrow parameter ranges. The distribution of parameter $P$ is rather scattered, but its peak point is very close to the actual value, and the same is true for $T_0$. This behaviour arises from parameter degeneracies inherent in RV orbital fitting, where $P$ and the reference epoch $T_0$ are strongly correlated (T. D. Brandt et al. 2021; C. Lam et al. 2024). Although the true values of mass ratio $q$ and primary mass $m_1$ fall within their respective posterior distributions, their uncertainty ranges are broader compared to those of $P$ or $e$. This phenomenon stems from the intrinsic degeneracy between $q$ and $m_1$ in radial velocity measurements. While we noticed that parametrizing the system using the radial velocity semi-amplitude ($K$) could reduce dimensionality and avoid this degeneracy, we chose to sample $q$, $m_1$, and $i$ directly to maintain physical interpretability. This approach allows for the incorporation of additional observational constraints, such as astrometry or photometry, where these parameters are directly relevant. Sampling $K$ would simplify the parameter space but limit the ability to explore the broader physical properties of the binary system.

The angle of inclination $i$, the angle of precession $\omega$, and the time of periastron passage $T_0$ are not intrinsic properties of binary orbits. These parameters depend on the relative geometric configuration between the binary system and the observer. Consequently, our primary focus is on the parameters $P$, $q$, $e$ and $m_1$. Fig. 1 demonstrates the feasibility of using DS of RVs to estimate the orbital parameters for a given binary system.

## 3 CUMULATIVE DISTRIBUTION CHARACTERISTICS OF DS OF RVS

The cumulative distribution curve (CDC, hereafter) encapsulates the distribution characteristics of a data set. Various attributes such as curve morphology, domain of definition, and slope at inflection points differ among distinct data sets. The utility of the proposed $\Delta$AJRV CDC for studying statistical properties of binary stars depends on understanding which factors influence curve characteristics and their underlying mechanisms. An observed sample set typically comprises both binary and single stars, with potentially heterogeneous spectral types among different samples. These samples exhibit variations in RV measurement errors, the number of repeated observations per source, the distribution functions of overall orbital parameters, and $f_{\mathrm{bin}}$.

Furthermore, a critical consideration is that for certain binary systems, radial velocity exhibits minimal temporal variation, with RV values predominantly or entirely falling within the range of measurement errors. This creates an ‘RV fuzzy interval’ that impedes the distinction between single and binary stars based solely on RV magnitude.

Consequently, we conducted systematic investigations across the following 5 distinct aspects to examine each in detail. Prior to conducting these experiments, certain conventions regarding symbols and assumptions must be established:

(i) The comprehensive collection of $\Delta$AJRV sequences from ALL the samples is designated as *DtajrvList*.

(ii) Each simulated group contained $10^5$ observational sources to construct the CDC. Each source featured 11 orbital samplings with random observation time interval within (0, $P$], resulting in a 10-element $\Delta$AJRV sequence per source.

(iii) We assumed that the orbital period $P$, mass ratio $q$, and eccentricity $e$ follow power-law distributions with exponents $\pi$, $\kappa$, and $\eta$, respectively:

$$f(P) \sim P^{\pi}, \quad f(q) \sim q^{\kappa}, \quad f(e) \sim e^{\eta} \tag{4}$$

(iv) The parameters $m_1$, $i$, $\omega$, and $T_0$ were assumed to follow uniform distributions. Parameter value ranges are presented in Table 2, with exceptions for the period $P \in [1, 1400]$ and inclination angle $i \in [-\pi/2, \pi/2]$. The minimum of $e$ was set to $10^{-5}$ to avoid the exception with a zero divisor (when $\eta < 0$) during the execution of the program code.

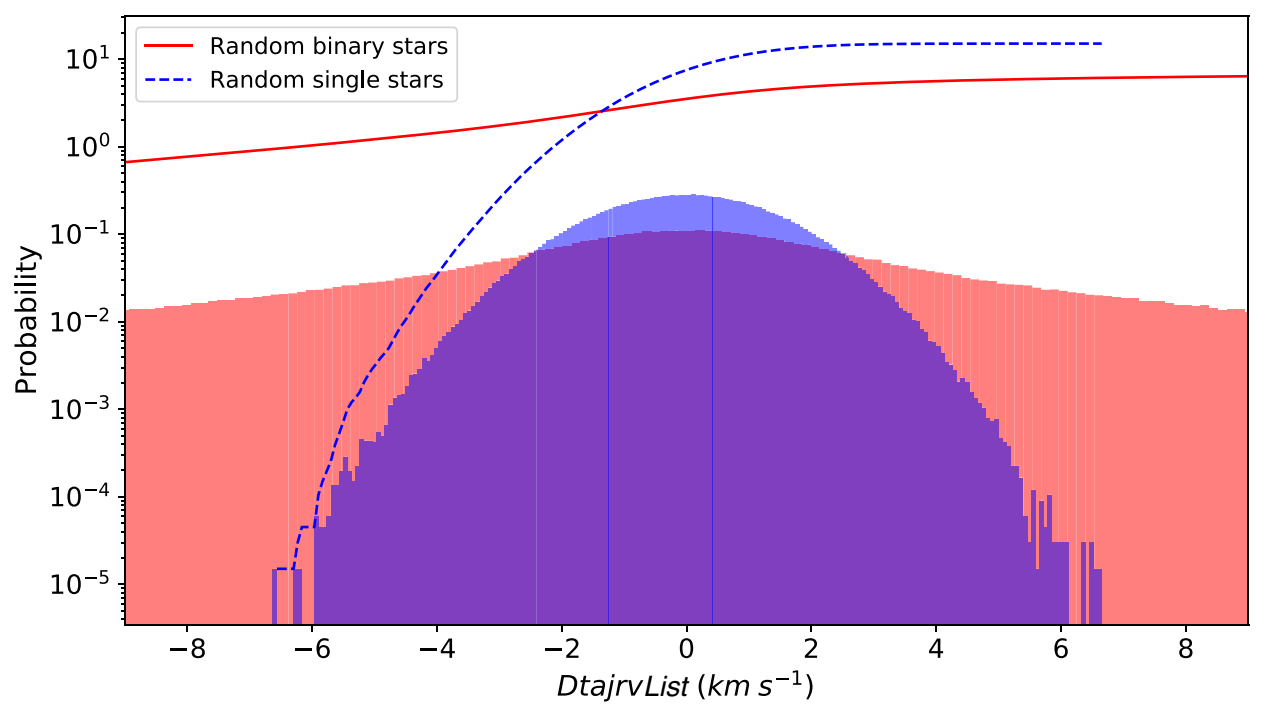

**Figure 2.** The distributions of $\Delta$AJRV for binary stars and single stars in the 'RV fuzzy interval'. The solid line represents binary stars, while the dotted line represents single stars.

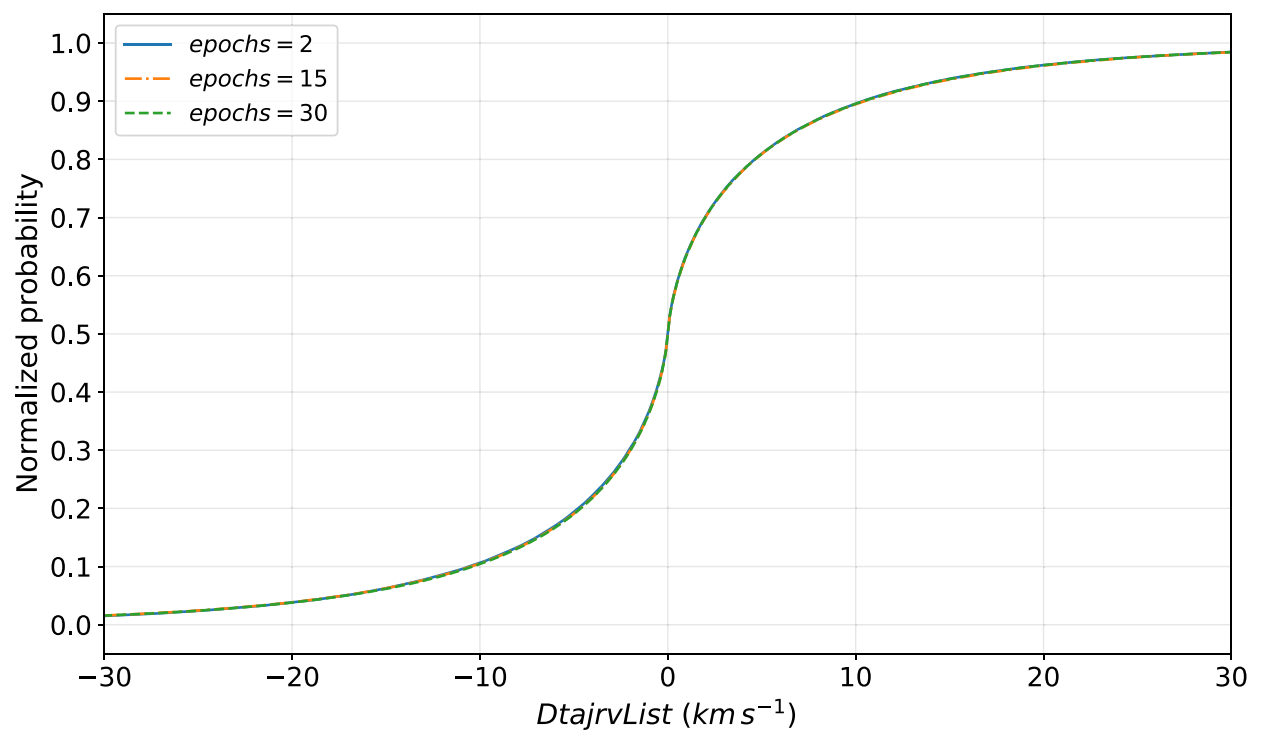

**Figure 3.** The CDCs of *DtajrvList* of binary-star samples with different epochs.

Then we can generate simulated orbits by randomly sampling these parameters.

### 3.1 The 'RV fuzzy interval' of single and binary stars

Actual observational data typically encompasses both binary systems and single stars. For certain binary systems with relatively extended orbital periods, radial velocity variations are minimal. Even if the orbital velocity changes significantly, if the orbital inclination $i$ is close to 0 or 180 degrees, the RVs can only show slight fluctuations. When the RV fluctuation is very small, the corresponding $\Delta$AJRV is also very small, which is very similar to that of single stars and therefore difficult to distinguish. In such circumstances, if one attempts to directly determine whether a particular observed source is a binary star or a single star using RV data alone, it would be extremely difficult, and the results would likely be inaccurate. It is worth noting that even for binary stars with relatively small $\Delta$AJRV, their measured RV not only include measurement errors but also contain components related to orbital motion. So, would their CDC exhibit differences?

We generated separate sets of binary-star and single-star samples, with the measurement error $\varepsilon$ set to 1.0 km s$^{-1}$. For the binary-star sample, we adopted $(\pi, \kappa, \eta) = (0, 0, 0)$, and radial velocities were calculated from equation (1). The radial velocities for single stars were sampled from a normal distribution $N(0, \varepsilon)$. The distributions of both are shown in Fig. 2. The red area in the figure represents binary stars, and the blue area represents single stars. Each CDC uses the same colour. It is evident that the cumulative distribution curves (CDCs) of binary stars and single stars exhibit significant differences. This experiment demonstrates that even within the 'RV fuzzy interval' where distinguishing between single and binary stars is challenging, sufficient sampling enables the identification of distinctive CDC patterns that differentiate these populations, analogous to distinct 'genetic signatures'. This characteristic enables the estimation of $f_{\rm bin}$ from mixed samples.

### 3.2 Number of observations

For single stars, regardless of the number of repeated observations, their simulated RVs are sampled from the distribution of $N(0, \varepsilon)$, and the CDC of $\Delta AJRV$ does not change. To investigate the CDC of binary stars under different numbers of observations, we conducted the following experiment: taking $(\pi, \kappa, \eta, \varepsilon) = (0, 0, 0, 1.0)$ and generated simulated samples with 2, 15, and 30 epochs, respectively. Apart from the number of observations, all other parameters are in accordance with the aforementioned conventions and assumptions. The CDC results are shown in Fig. 3.

$$N(0, \varepsilon) \tag{10}$$

The spacing between each point on the horizontal axis in the figure is 0.1, which is 1/10 of $\varepsilon$. From the figure, it is impossible to discern any difference with the naked eye. Therefore, we compute the mean absolute residual across the three learning curves. The corresponding values are $1.346 \times 10^{-5}$ (between 2- and 15-epoch samples), $2.8 \times 10^{-7}$ (between 2- and 30-epoch samples), $1.374 \times 10^{-5}$ (between 15- and 30-epoch samples).

This fact implies that $\Delta$AJRV shows no significant variation with the number of observations in pure binary-star samples with the same orbital parameter distribution laws. Indeed, changing the cadence will sample different orbital phases for different binaries, but it does not change the underlying distribution of $\Delta$AJRV: in the limit of many targets, the CDCs converge for any reasonable choice of epochs. The $\Delta$AJRVs obtained from samples with 2 epochs can be regarded as a subset of those from 15 epochs, and likewise the 15-epoch set is a subset of the 30-epoch set; they all arise from the same 'background' distribution. When the sample size is relatively small, CDCs can differ due to statistical fluctuations, but these fluctuations diminish as the sample grows, leading to near-identical CDCs. In other words, when the sample size is large, the distribution of $\Delta$AJRV is not sensitive to the number of observations.

This characteristic enables us to construct the 'basic data template' for comparison without being limited by the number of observations of the sample. It also means that our 'template' can have greater universality in terms of the number of observations. In DVCD we generate $10^5$ observation sources per hyperparameter combination $(\pi, \kappa, \eta, \varepsilon)$ to ensure sufficient sampling and robust statistical stability.

### 3.3 The measurement error of RVs

The measurement error $\varepsilon$ of RV is an important influencing factor. As the standard deviation of a normal distribution, it directly determines the scale of the $\Delta$AJRV distribution for both single and binary stars, and undoubtedly has a significant impact on the CDC of the *DtajrvList*. Let $(\pi, \kappa, \eta) = (0, 0, 0)$, the number of observations be 11, and $\varepsilon = (0.1, 0.5, 1, 2, 4)$. Generate samples

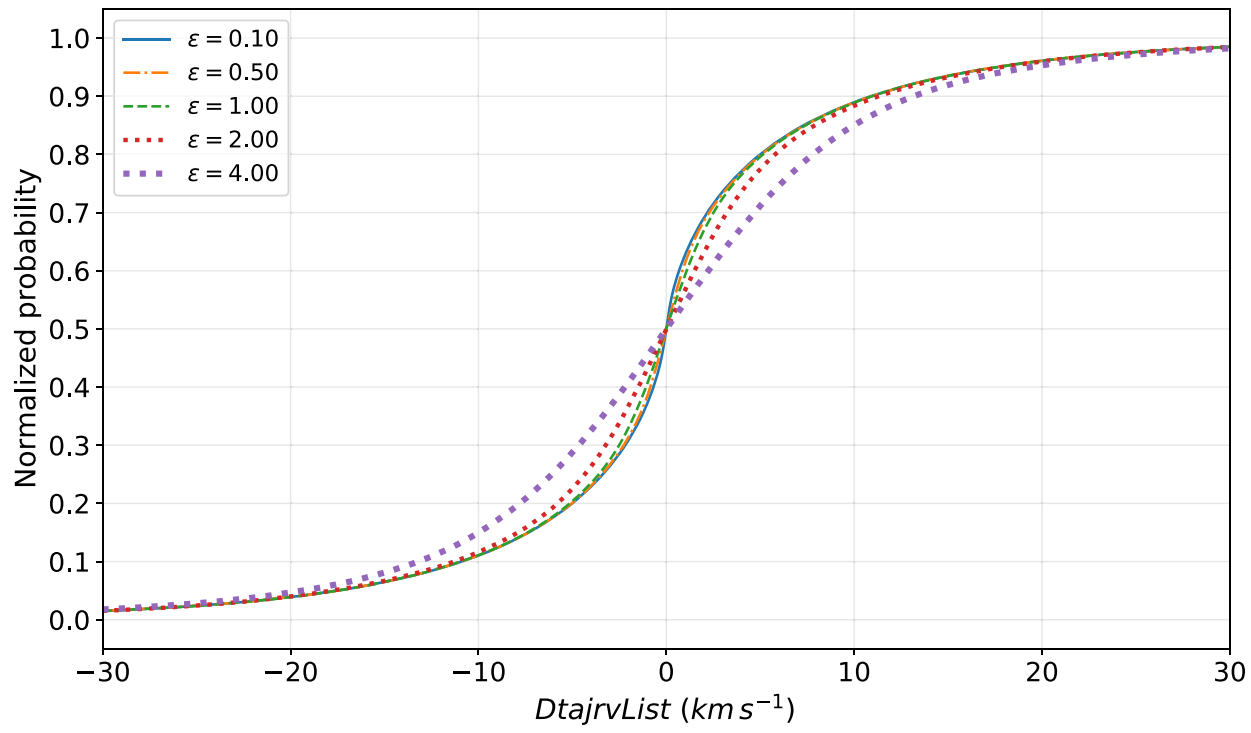

**Figure 4.** The CDCs of *DtajrvList* of binary-star samples with different $\varepsilon$.

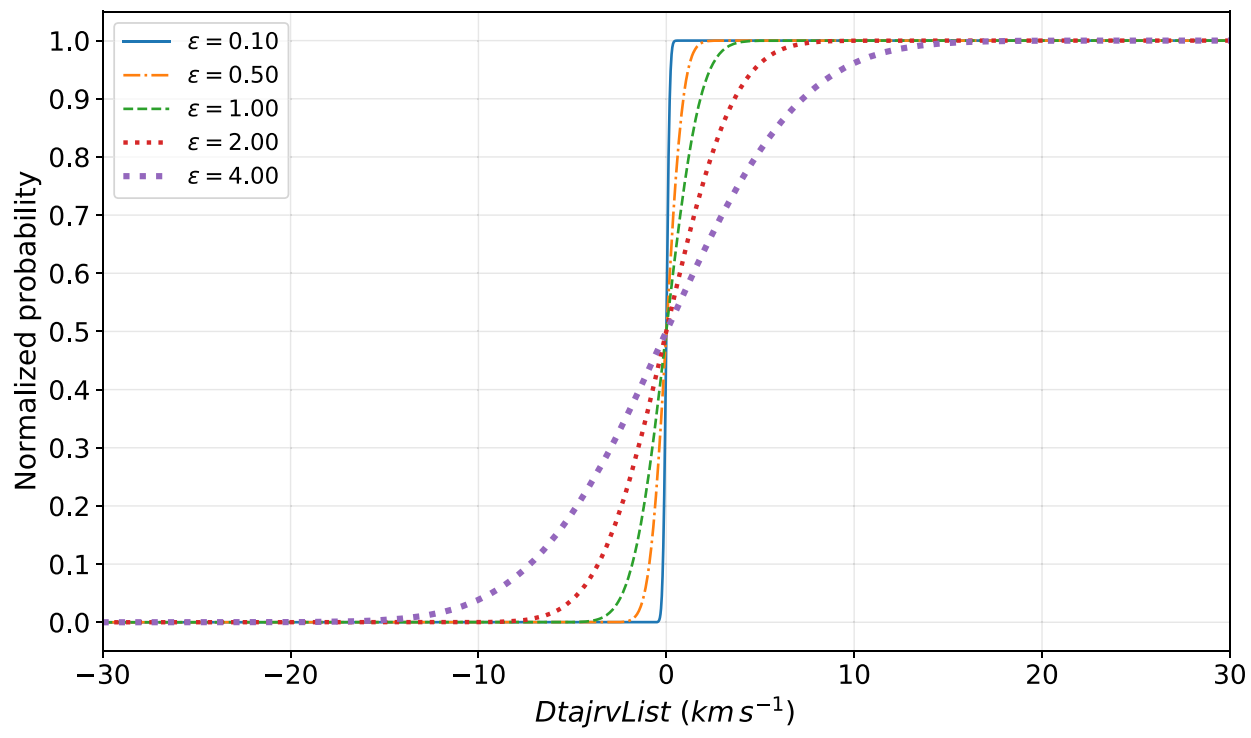

**Figure 5.** The CDCs of *DtajrvList* of single-star samples with different $\varepsilon$.

of the same size for both single- and binary-star samples, respectively, as shown in Figs 4 and 5.

It can be seen that as $\varepsilon$ increases, the dispersion of the *DtajrvList* for the binary-star system becomes larger, and the CDC becomes more gradual. The situation for the single star is similar, except that the degree of change in CDC are stronger than that of binaries. Such results are consistent with the theoretical analysis. This indicates that in terms of RV measurement errors, this algorithm is consistent with other algorithms. The value of $\varepsilon$ is a sensitive factor for the accuracy of the final calculation result.

### 3.4 Distribution laws of orbital parameters

In the case of multiple samples, the orbital period $P$, mass ratio $q$, and eccentricity $e$ of binary systems may follow certain distribution laws. Previous studies have investigated the distributions of these orbital parameters and proposed different hypotheses, such as assuming that the orbital period $P$ follows either a power-law (H. Sana et al. 2013; G. Yanjun et al. 2022) or log-normal (D. Raghavan et al. 2010) distribution. For unknown distribution laws, we consider assuming power-law prior distributions for these parameters. This approach is advantageous because, given identical parameter domains, negative power exponents correspond to decreasing functions while positive exponents correspond to increasing functions. Estimating the power exponents provides general insights into the increasing or decreasing trends of parameter distributions.

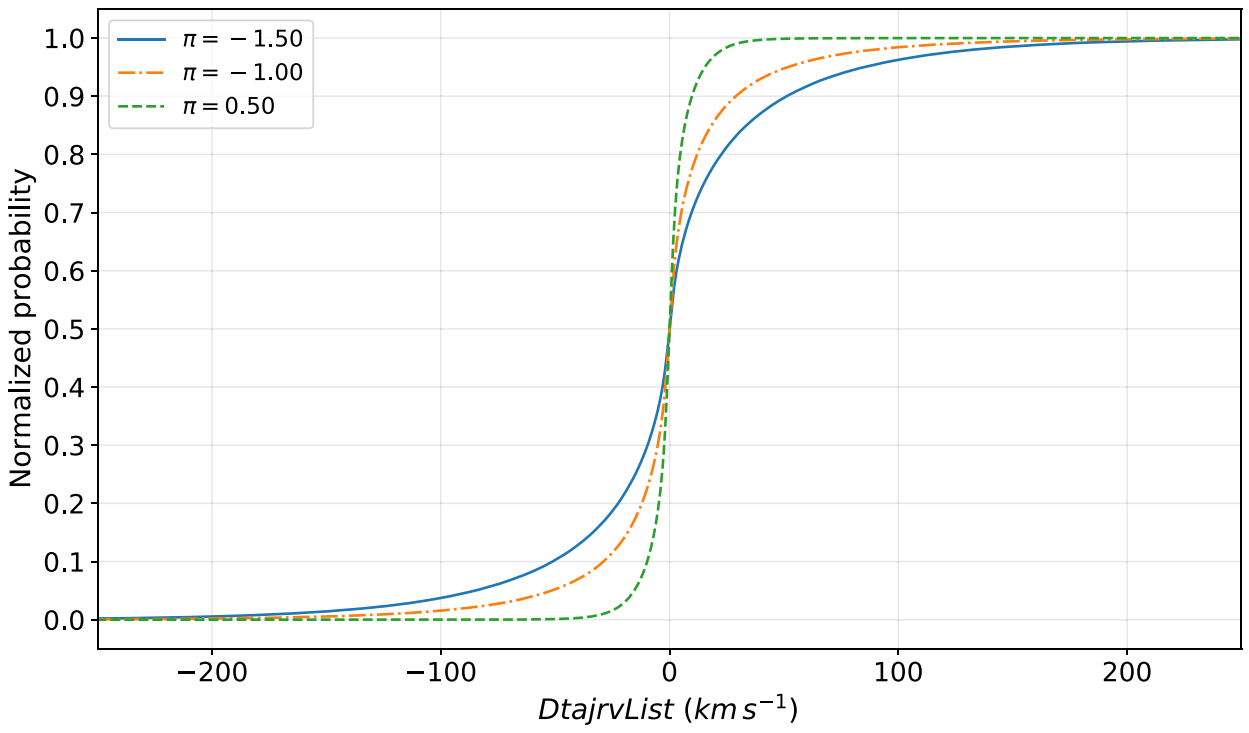

**Figure 6.** The cumulative curves of *DtajrvList* with different $\pi$.

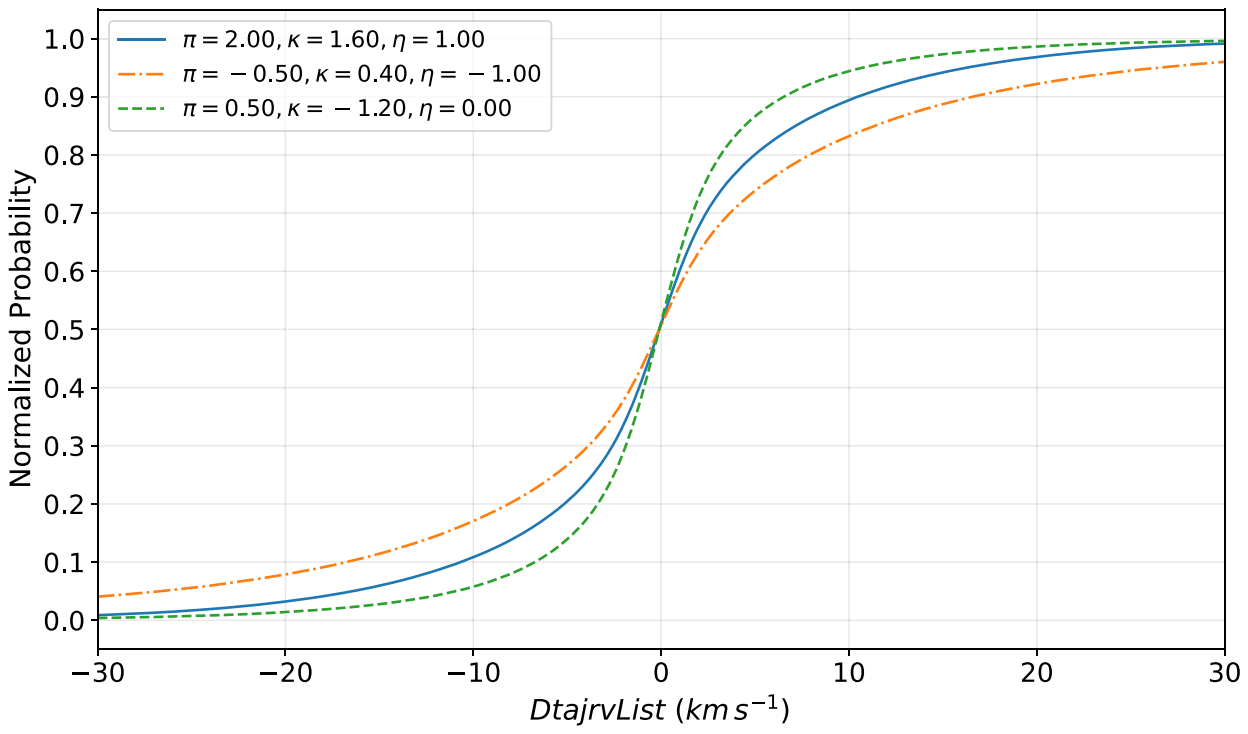

**Figure 7.** The difference of the cumulative distribution curves of *DtajrvList* of binary-star samples generated with different $\pi$, $\kappa$ and $\eta$.

As evident from equation (1), the radial velocity equation incorporates seven fundamental parameters. When including observation time, eight quantities are required to determine a radial velocity measurement. Consequently, the prior distribution laws governing these parameters significantly influence the final radial velocity distributions. For instance, an increasing probability density function for orbital period $P$ implies fewer binary systems with short periods and more systems with longer periods. In such cases, the peak or median of the radial velocity distribution shifts towards smaller values. Conversely, a decreasing probability density function for $P$ produces the opposite effect, shifting the distribution towards larger values. These variations correspondingly affect the cumulative distribution characteristics of *DtajrvList*.

To verify the above-mentioned theoretical analysis, we first examined the influence of $\pi$ on *DtajrvList*. Let $\pi$ be equal to $(-1.5, -1, 0.5)$, $(\kappa, \eta, \varepsilon) = (0, 0, 1.0)$, and $epochs = 11$. We generated simulated data and observed the CDC of *DtajrvList*, as shown in Fig. 6. It can be seen that, under the condition that other factors remain unchanged, different prior distributions of $\pi$ have an obvious impact on the final distribution of *DtajrvList*. It is worth noting that near the 'RV fuzzy interval' the differences between different curves are visible to the naked eye.

Next, it is necessary to examine the CDC situation of the combination $(\pi, \kappa, \eta)$ when they take different values. By assigning different values to $(\pi, \kappa, \eta)$, the CDC of them still shows significant differences in the 'RV fuzzy interval', as shown in Fig. 7. This characteristic meets our need to distinguish different distribution laws of orbital parameters through CDC.

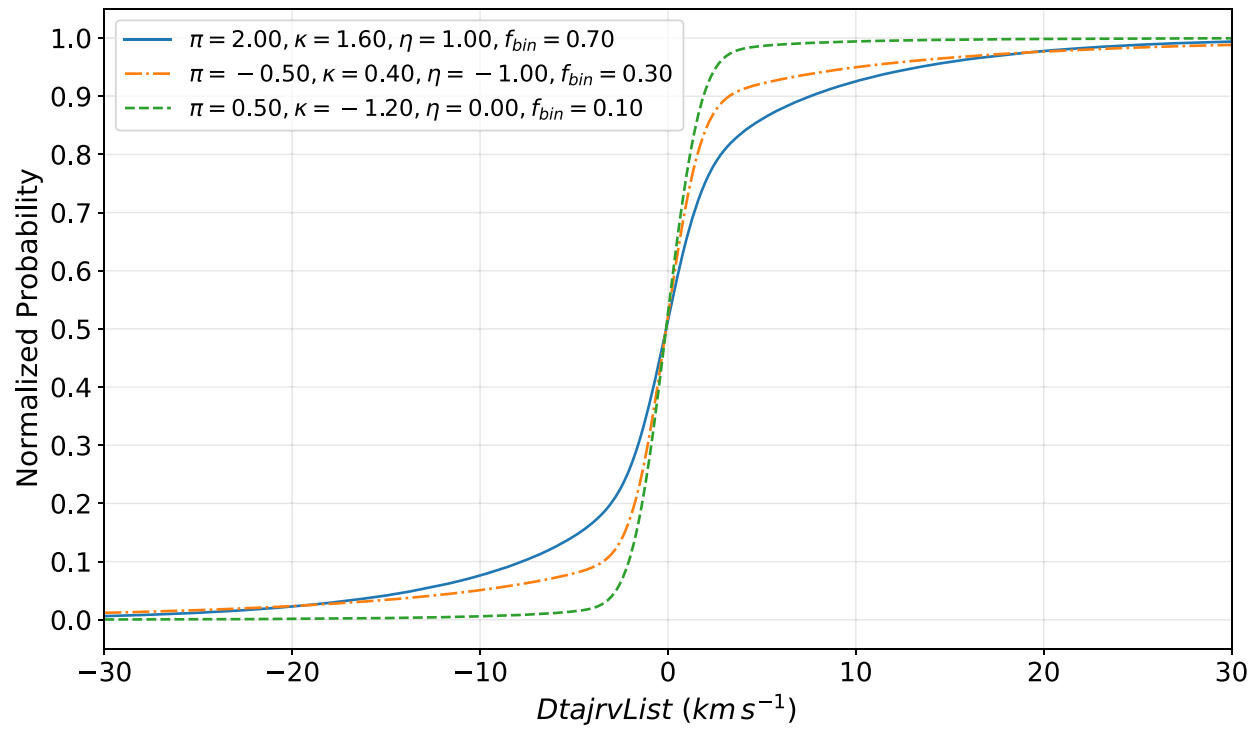

**Figure 8.** The difference of the cumulative distribution curves of *DtajrvList* of binary-star samples generated with different $(\pi, \kappa, \eta)$ and $f_{\rm bin}$.

### 3.5 The ground truth of binary fraction

According to the analysis in the previous section, the CDCs under different distribution laws of orbital parameters will show significant differences. Then, for samples with different values of $(\pi, \kappa, \eta)$ and $f_{\rm bin}$, their CDC may also be different. Based on Fig. 7, we further mixed in single-star samples to generate simulated data with $f_{\rm bin}$ equal to 0.7, 0.3, and 0.1, respectively. The CDC results are shown in Fig. 8. From Figs 7 and 8, it can be found that the cumulative distribution of the *DtajrvList* varies under different $(\pi, \kappa, \eta)$ and $f_{\rm bin}$.

The above-mentioned experiments show that first, within the 'RV fuzzy interval', the cumulative distribution characteristics of *DtajrvList* can be clearly distinguished under different prior conditions. Then, outside the 'RV fuzzy interval', the CDCs of *DtajrvList* for single and binary stars are obvious different; the CDC for pure binary stars under different orbital parameter distribution laws also shows more or less differences. These properties are very useful for us to constrain the orbital parameters' distribution laws and the $f_{\rm bin}$. Through experimental verification, the parameters we want to predict or estimate will all have different effects on the CDC of *DtajrvList*, showing different characteristics. This provides an experimental basis for us to further develop algorithms.

## 4 THE DVCD ALGORITHM

### 4.1 Implementation

The core challenge in algorithmic implementation is to accurately determine whether two samples originate from the same underlying distribution, or equivalently, to quantify the statistical similarity between their distributions. To achieve this, the two-sample Kolmogorov–Smirnov (KS) test is utilized. This test compares the empirical cumulative distribution functions (ECDFs) of the two samples to assess their agreement. Importantly, the KS test produces a $p$-value that measures the strength of evidence against the null hypothesis, which assumes that the two samples are drawn from the same continuous distribution. A higher $p$-value indicates greater consistency with the null hypothesis, suggesting stronger compatibility between the distributions of the two samples.

In practice, generating fully sampled simulated data sets on-the-fly for each evaluation is computationally prohibitive. To address this, we precompute a high-statistics template library over a systematic grid of hyperparameters – including orbital parameter-distribution exponents $(\pi, \kappa, \eta)$ and assumed measurement errors $(\varepsilon)$ – and compare observed $\Delta$AJRV cumulative distributions against convex combinations of these templates. This 'space-for-time' strategy trades storage overhead for substantial runtime reduction while preserving the statistical fidelity essential for robust KS-based inference.

**Table 3.** Hyperparameter grid structure.

| Name | Range | Step | Points |
|---|---|---|---|
| $\pi$ | $[-2,\ 2]$ | 0.1 | 41 |
| $\kappa$ | $[-2,\ 2]$ | 0.1 | 41 |
| $\eta$ | $[-2,\ 2]$ | 0.1 | 41 |
| $\varepsilon$ | 1.0 | – | 1 |

Specifically, for each hyperparameter combination $(\pi, \kappa, \eta, \varepsilon)$, we simulate $10^5$ independent orbital realizations; for each orbit, we draw 11 independent RV measurements – yielding $10^6$ $\Delta$AJRV values per base template. This design ensures both statistical sufficiency and sampling thoroughness, while eliminating redundant recomputation across multiple downstream analyses. During comparison, a synthetic mixed $DtajrvList_{\rm tmpl}$ is constructed by drawing $\Delta$AJRV values from the binary-star and single-star template libraries in proportions $f_{\rm bin}$ and $1 - f_{\rm bin}$, respectively. The $f_{\rm bin}$ is systematically varied from 0.0 to 1.0 in increments of 0.01, and the corresponding KS $p$-value is computed at each step (cf. equation 5). By analysing the resulting $p$-value surface across the $f_{\rm bin}$-parameter space, we infer binary population characteristics with statistical rigour and interpretability. These operations can be summarized as the following steps:

(i) **Precompute templates:** Generate a high-statistics library of templates for a grid of hyperparameters $(\pi, \kappa, \eta, \varepsilon)$, simulating $10^5$ orbits per combination and storing $10^6$ $\Delta$AJRV values per template.

(ii) **Construct mixed samples:** Create synthetic $\Delta$AJRV samples by combining binary-star and single-star templates in proportions $f_{\rm bin}$ and $1 - f_{\rm bin}$, respectively.

(iii) **Perform KS test:** Use the KS test to compare observed $\Delta$AJRV distributions with mixed samples, calculating $p$-values to assess distributional similarity.

(iv) **Optimize binary fraction:** Systematically vary $f_{\rm bin}$ from 0.0 to 1.0 in increments of 0.01, identifying the value that maximizes the KS $p$-value.

(v) **Infer population characteristics:** Analyse the $p$-value surface across the parameter space to estimate binary population properties with statistical rigour:

$$\begin{aligned} p\text{–value}\ (\pi, \kappa, \eta, \varepsilon, f_{\rm bin}) \\ = P_{\rm KS}\ (DtajrvList_{\rm obs}, DtajrvList_{\rm tmpl}(\pi, \kappa, \eta, \varepsilon, f_{\rm bin})). \end{aligned} \quad (5)$$

### 4.2 Validations with mock data

A hyperparameter grid was systematically constructed in accordance with the specifications outlined in Table 3. At each grid point (combination of hyperparameters), a 'basic template' was generated and persistently stored, yielding a total of 68 921 templates (41 × 41 × 41 × 1). This precomputed template set is reused across all three subsequent experiments – loaded from storage rather than regenerated in real time – thereby eliminating redundant computation and substantially enhancing experimental throughput. Furthermore, this template set remains valid for all samples characterized by an RV measurement error of $1.0\,km\,s^{-1}$, ensuring consistent and efficient reuse without loss of fidelity.

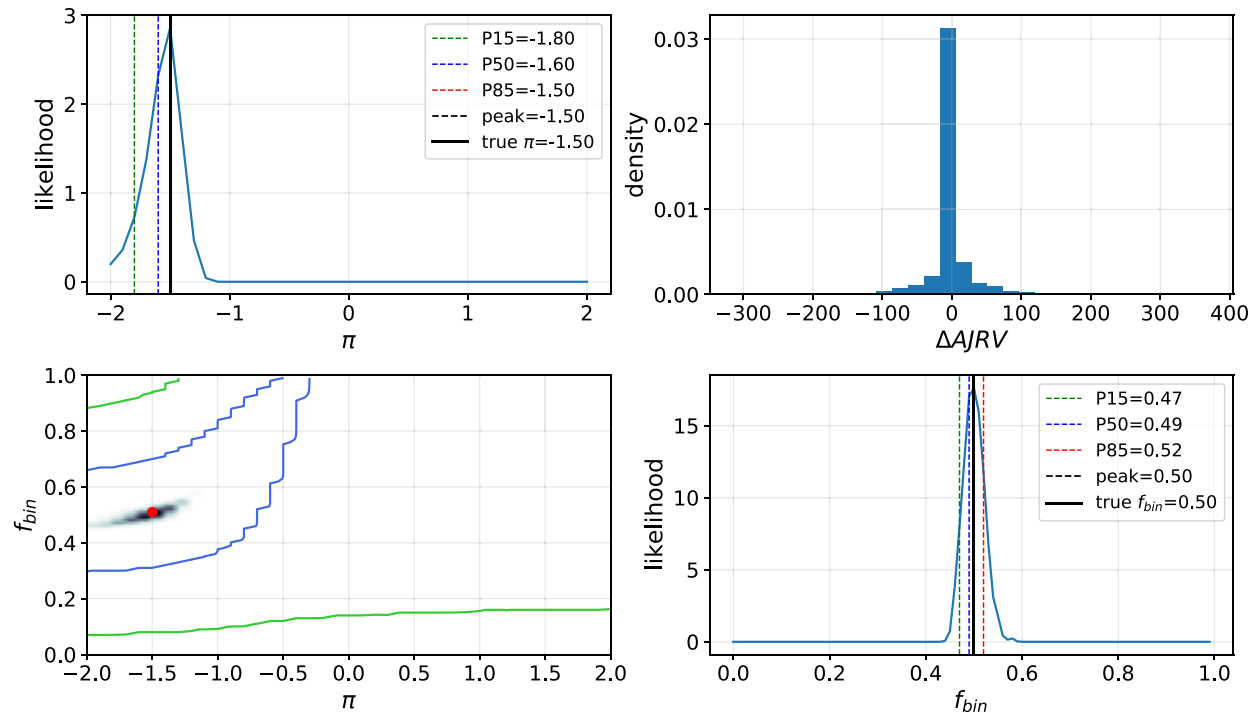

**Figure 9.** Calculated results of samples whose $\pi$ equals $-1.5$, the black solid lines in the graph represent the position of true values. The small dot in the lower left sub-image indicated the peak position.

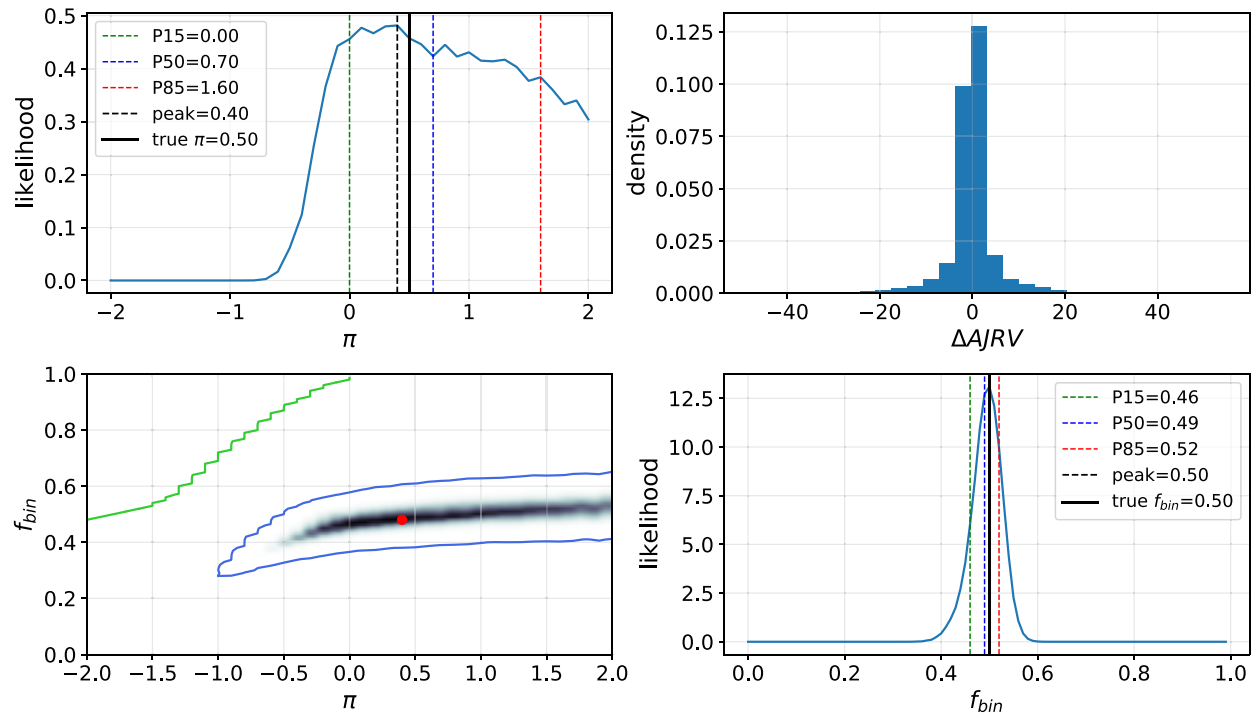

**Figure 10.** Calculated results of samples whose $\pi$ equals 0.5, the black solid lines in the graph represent the position of true values.The small dot in the lower left sub-image indicated the peak position.

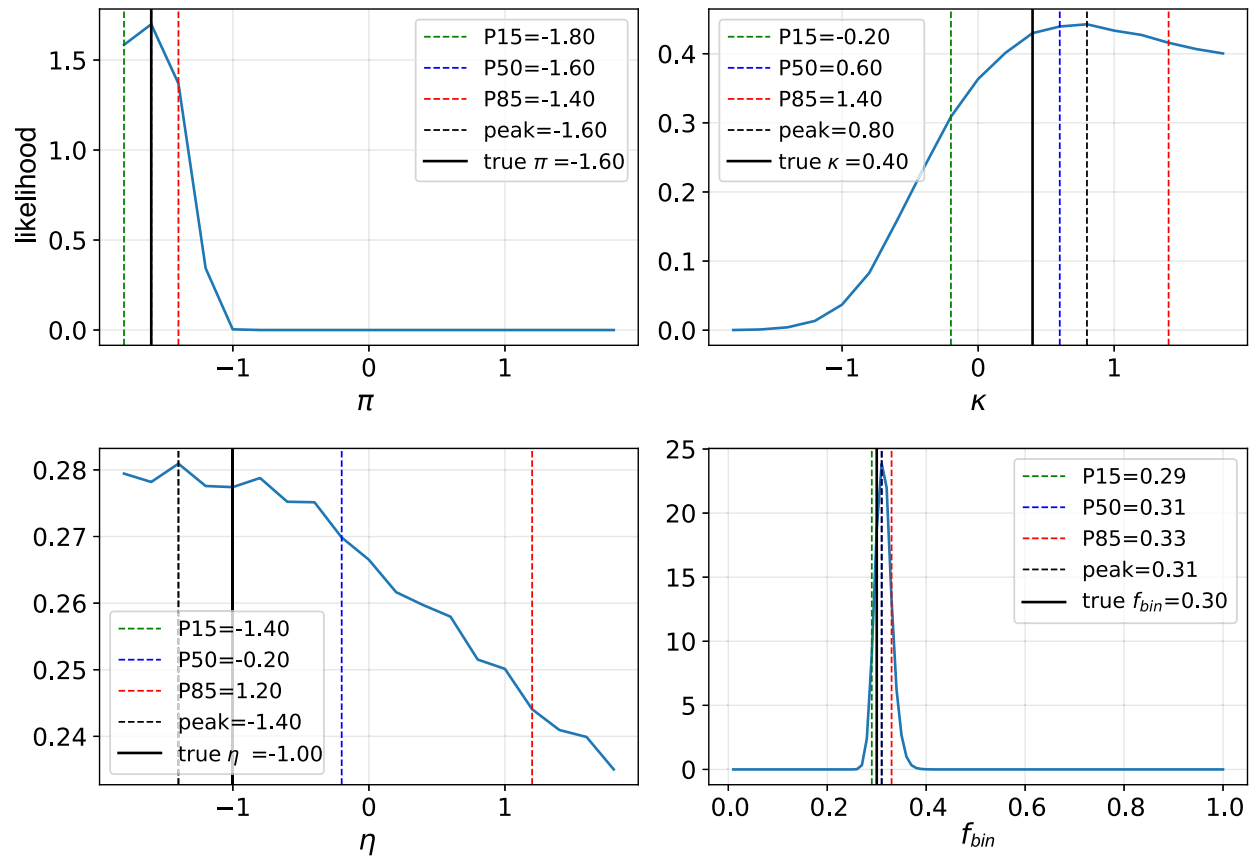

**Figure 11.** The difference of the cumulative distribution curves of *DtajrvList* of binary-star samples generated with different $\pi$, $\kappa$, $\eta$ and $f_{\mathrm{bin}}$.

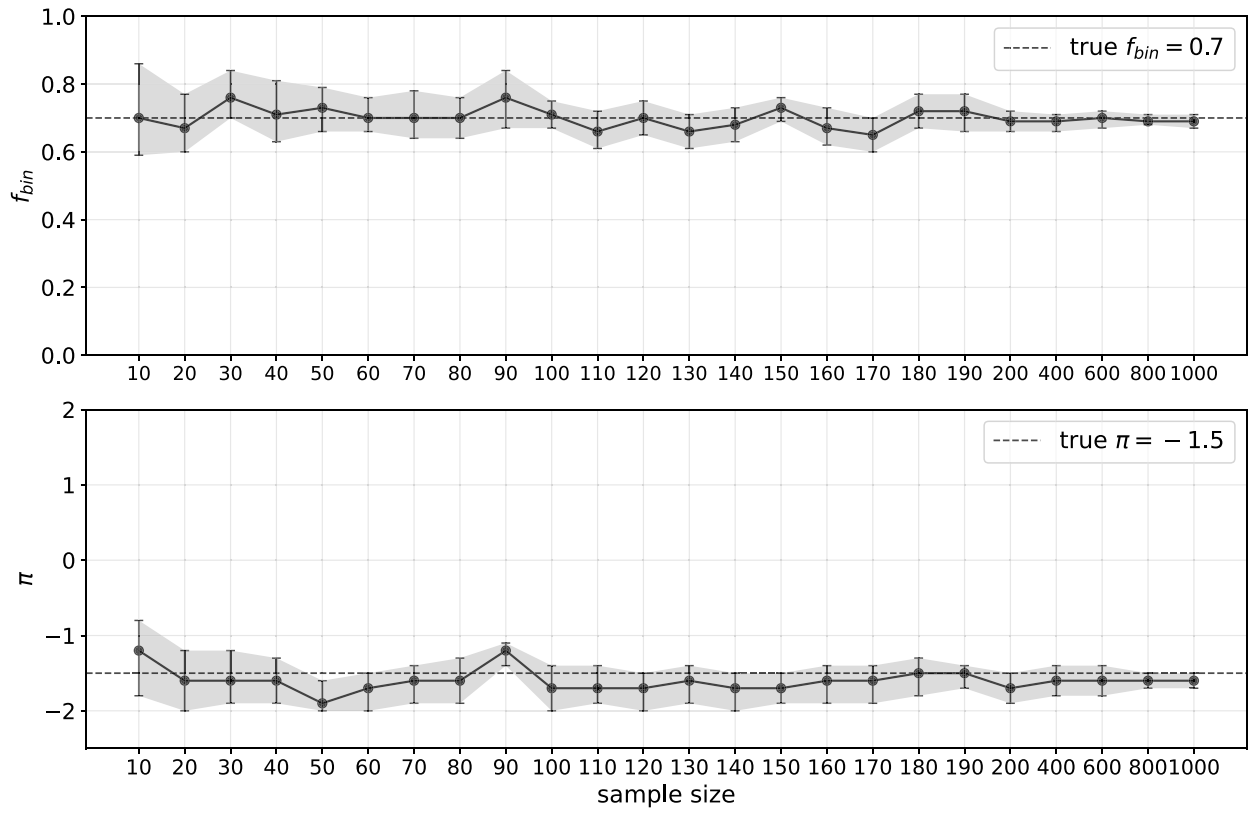

**Figure 12.** The predictions of $\pi$ and $f_{\mathrm{bin}}$ with different sample sizes.

Two groups of mock samples were generated with a sample size equal to 1000, epochs 11, $\varepsilon = 1.0$ km s$^{-1}$, true $(\kappa, \eta, f_{\mathrm{bin}}) = (0, 0, 0.5)$ but different $\pi = (-1.5, 0.5)$. The ranges of values for the orbital parameters other than $P$ and $e$ are shown in Table 2. The upper limit of $P$ was set to 3000 d to contain systems with longer orbital periods. In the subplots of the following three figures, 'likelihood' denotes the result derived from the normalization of the $p$-value and its subsequent conversion into a probability, such that the area under the curve sums to 1.

The calculated results using the DVCD algorithm of the two groups are shown in Figs 9 and 10, respectively. From the two graphics, we can see that the truths of both $\pi$ and $f_{\mathrm{bin}}$ can be restored accurately.

Furthermore, we generate a group of samples with true $(\pi, \kappa, \eta, f_{\mathrm{bin}}) = (-1.6,\ 0.8,\ -1.4,\ 0.3)$ to test if the truths can be restored. The results are displayed in Fig. 11. It can be seen that $\pi$ and $f_{\mathrm{bin}}$ were well constrained, while the error range of $\kappa$ and $\eta$ is wider than that of $\pi$. Two reasons may cause this phenomenon. On the one hand, the sample size 1000 and true $f_{\mathrm{bin}} = 0.3$ mean that there are few binaries that carry orbital information to predict the distribution laws of $\kappa$ and $\eta$. On the other hand, the prior distributions of $q$ and $e$ do not significantly affect *DtajrvList* more than $\pi$ and $f_{\mathrm{bin}}$. Generally speaking, the accuracy of the estimates of $f_{\mathrm{bin}}$ was better than those of $\pi$, $\kappa$, and $\eta$.

## 4.3 The influencing factors of calculation results

For simplicity, we generate samples with fixed $(\pi, \kappa, \eta) = (-1.5, 0, 0)$, $\varepsilon = 1.0$, and examine the variations of the calculation results with sample size, the observation epochs, and the true value of $f_{\mathrm{bin}}$.

### 4.3.1 *Sample sizes*

24 sets of simulated data with 16 observational epochs, $f_{\mathrm{bin}} = 0.7$, were generated with the sample size gradually increasing. The calculation results of the DVCD algorithm are shown in Fig. 12. Black dots between each of the error bars indicate the 50th percentile of the $p$-value, the lower and upper limits represent the 15th and 85th percentiles respectively, the same below.

First, the result of $f_{\mathrm{bin}}$. As shown in the graph, even when the sample size is very small (less than 100), the DVCD algorithm can accurately reproduce the true value of $f_{\mathrm{bin}}$. As the sample size increases, the 50th percentile of $f_{\mathrm{bin}}$ fluctuates around the true value, and the error gradually decreases. When the sample size reaches 200 or above, the error bars further shorten. Moreover, from an overall perspective, the error bars of all 24 sets of data cover the true value, indicating that the DVCD algorithm has a higher accuracy in estimating the $f_{\rm bin}$.

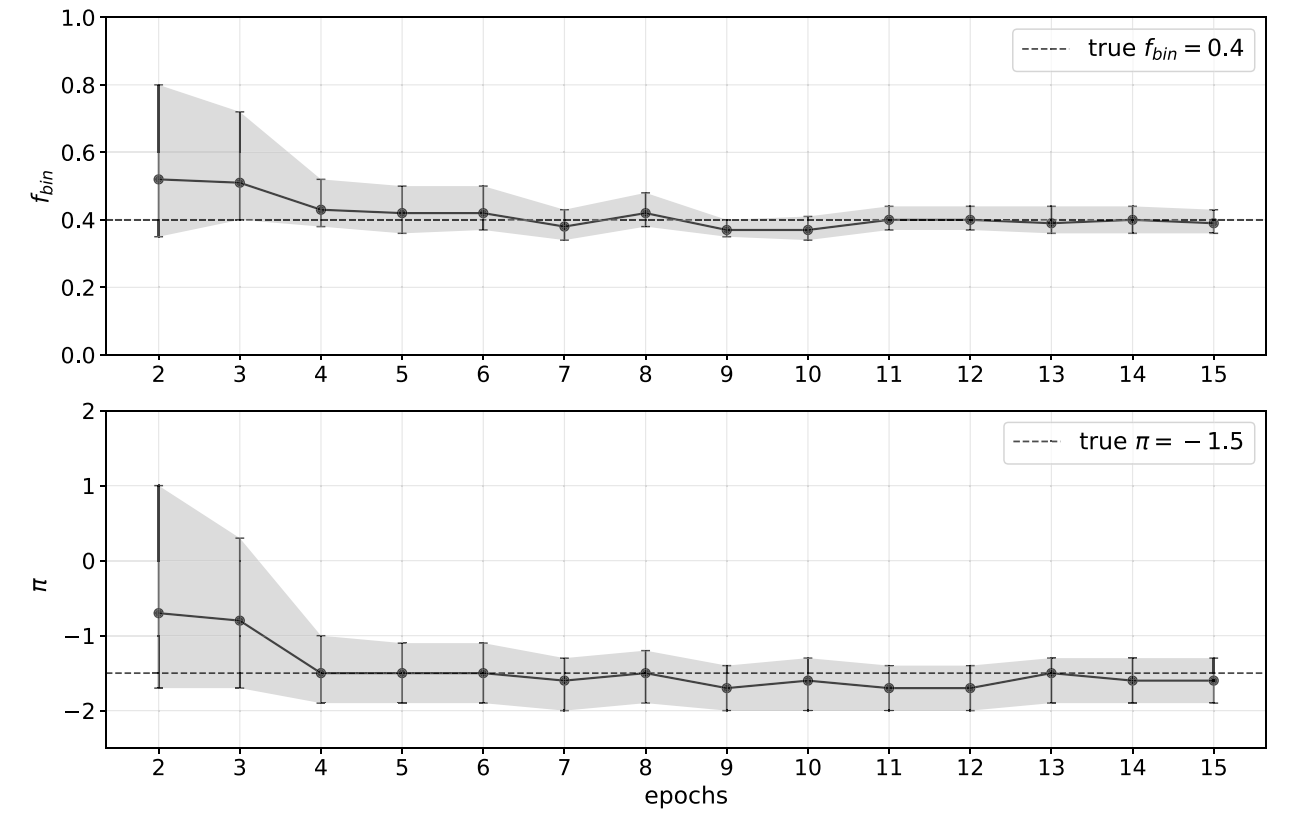

**Figure 13.** The predictions of $\pi$ and $f_{\rm bin}$ with different observational epochs.

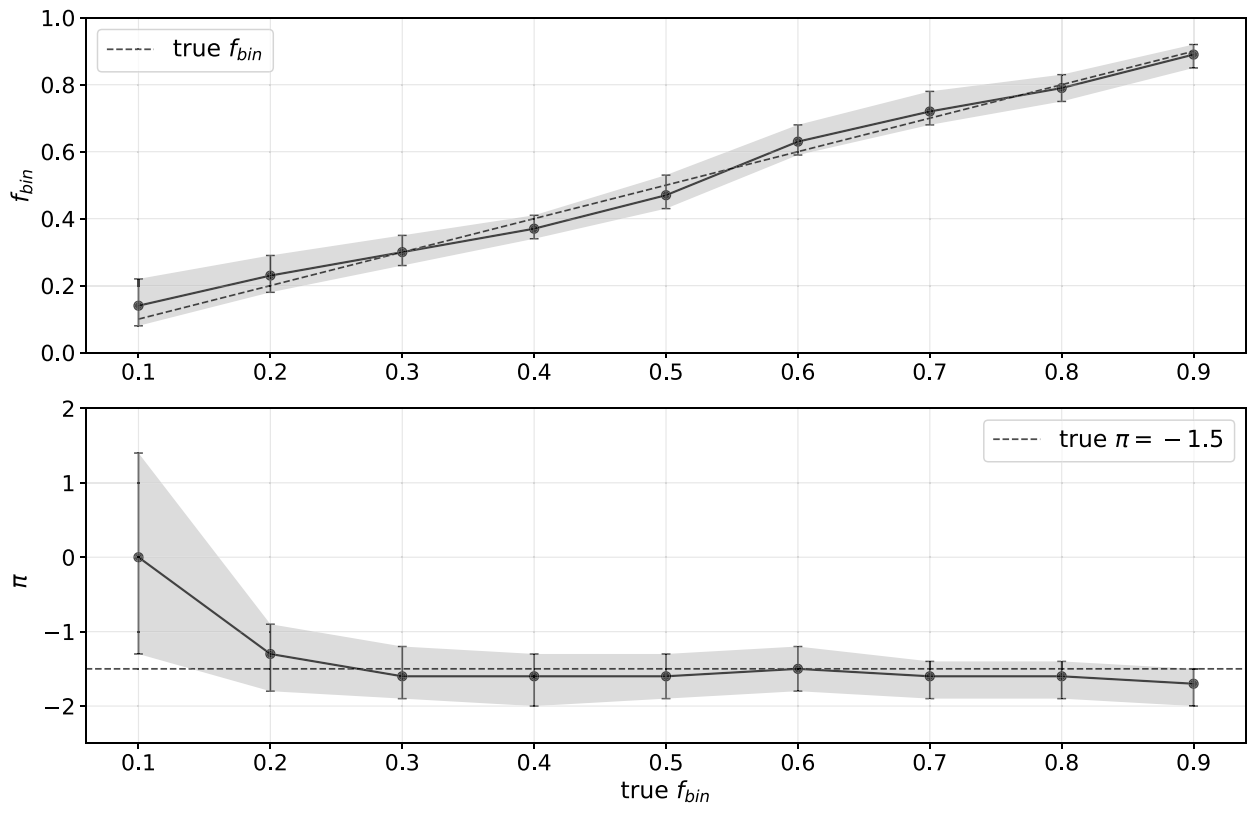

**Figure 14.** The predictions of $\pi$ and $f_{\rm bin}$ with different truths of $f_{\rm bin}$.

Second, the result of $\pi$. As can be seen from the figure, when the sample size is small, the 50th percentile fluctuates near the true value and the error bars are large. Among them, when the sample size is 90, the calculated result fails to cover the true value within the error range. This might be caused by the random processes in generating the data. When the sample size reaches 100 or more, all the prediction results successfully cover the true values. At the same time, overall, the error bars show a trend of gradually shortening. This indicates that if one wants to obtain reliable predictions of the parameter distribution laws, the sample size should be as large as possible.

### 4.3.2 *Epochs*

The simulated samples used in this experiment had an $f_{\rm bin} = 0.4$, sample size of 200, and the range of epochs was from 2 to 15, totalling 14 groups.

First, the result of $f_{\rm bin}$. As shown in Fig. 13, careful observation reveals that as the number of observations increases from 2 to 7, the error gradually decreases. When it reaches 8 or more, the error bars significantly shorten. This is in line with the intuitive feeling that the more observations, the more accurate the result. Because the more observations there are, the more obvious the distribution characteristics of $\Delta$AJRV brought by the orbital motion of the binary systems, making it easier to distinguish it from the single star within the 'RV fuzzy interval'.

Second, the result of $\pi$. When the number of observations is 2, the calculated result deviates significantly from the true value. This is not difficult to understand since the sample with only 2 observations provides us with very little information about the orbital information. When the epochs are equal to 3, the calculated value begins to converge towards the true value. When it reaches 7 or more, the predicted result is highly consistent with the true value, and the error bars narrow simultaneously. From the overall trend, it is undoubtedly true that the more observations there are, the more reliable the prediction results will be.

### 4.3.3 *Truths of* $f_{\rm bin}$

The simulation samples with a sample size of 200 and 10 epochs were generated, with the $f_{\rm bin}$ ranging from 0.1 to 0.9. We calculated these 9 sets of data using the DVCD algorithm, and the results are shown in Fig. 14.

First, the results of $f_{\rm bin}$. As can be seen from the graph, the 50th percentile of the *p*-value has always remained close to the ground truth, and the true values all fall within the error range. Especially when $f_{\rm bin}$ is less than or equal to 0.3, the error range of this algorithm is still very small.

Second, the results of $\pi$. As shown in the figure, when the true value of $f_{\rm bin}$ is 0.1, the error range of $\pi$ is relatively large. Starting from $f_{\rm bin} = 0.2$ the results move closer to the true values, the error bars also shrink significantly. As the $f_{\rm bin}$ increases, the error bars continue to shorten, indicating that the accuracy of the prediction results is getting better and better. This finding emphasizes that when analysing data sets with relatively small sample sizes and obtaining low $f_{\rm bin}$, the accuracy of orbital parameter distribution estimates may be compromised. Specifically, for $f_{\rm bin}$ values of 0.1 or 0.2, the predicted orbital parameter distributions may exhibit significant deviations from the true values. In such cases, the results should be interpreted with appropriate caution.

## 5 SCIENTIFIC APPLICATION

### 5.1 Red giant samples from APOGEE DR16

We used observational data of red giant stars selected from APOGEE DR16 to test the DVCD algorithm. C. Liu et al. (2014) presents a simple strategy for preliminary selection of red giants by $T_{\rm eff}$ and $\log g$ (see the green polygon shown in Fig. 15). The regions are considered to have better red giant completeness where $\log g < 4$ when $4600 < T_{\rm eff} < 5600$ or $\log g < 3.5$ when $T_{\rm eff} < 4600$. Note that the APOGEE catalogue includes spectra with bad pixels or low SNR. Therefore, we should pay attention to the markers in the catalogue (N. W. Troup et al. 2016; A. M. Price-Whelan et al. 2018). Our data filtering criteria mainly included the following:

(i) SNR $\geq 20$;
(ii) There are at least 6 repeated observations;
(iii) $\log g < 4$ when $4600 < T_{\rm eff} < 5600$ or $\log g < 3.5$ when $T_{\rm eff} < 4600$;

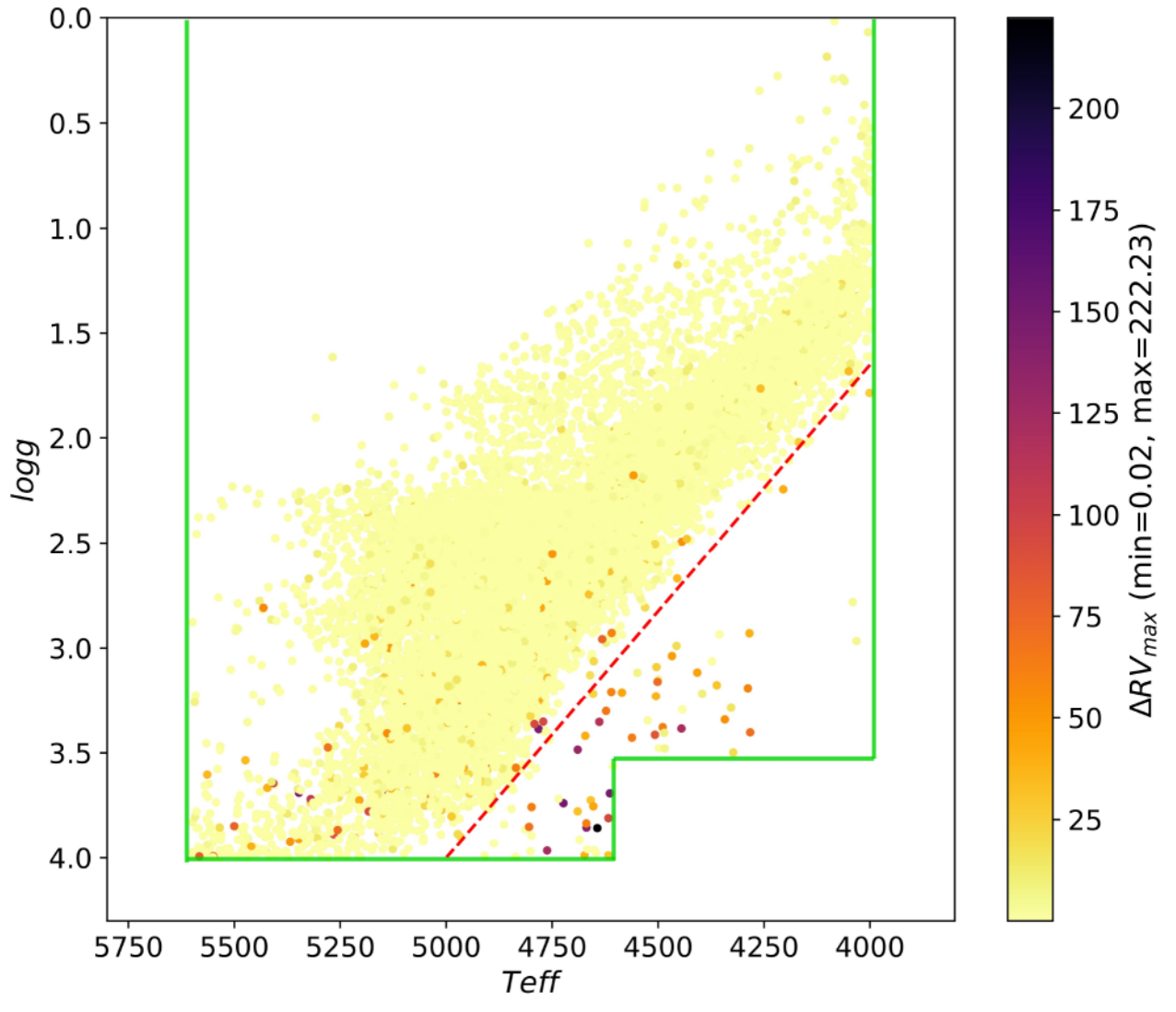

**Figure 15.** Red giant samples from APOGEE DR16.

(iv) Remove spectra satisfying: In 'APOGEE_STARFLAG' field, the 'LOW_SNR', 'VERY_BRIGHT_NEIGHBOR', 'PERSIST_JUMP_NEG', 'PERSIST_JUMP_POS', 'BAD_PIXELS', and 'PERSIST_HIGH' markers are equal to 1. In 'APOGEE_ASPCAPFLAG' field the 'STAR_BAD' marker is equal to 1.

A total of 15 182 sources were sought out, then we eliminated some abnormal data at the lower right dotted line in Fig. 15 and 15 112 were left. The basic information about them was depicted in Fig. 16. Where the observational intervals are time intervals between any pairs of observations.

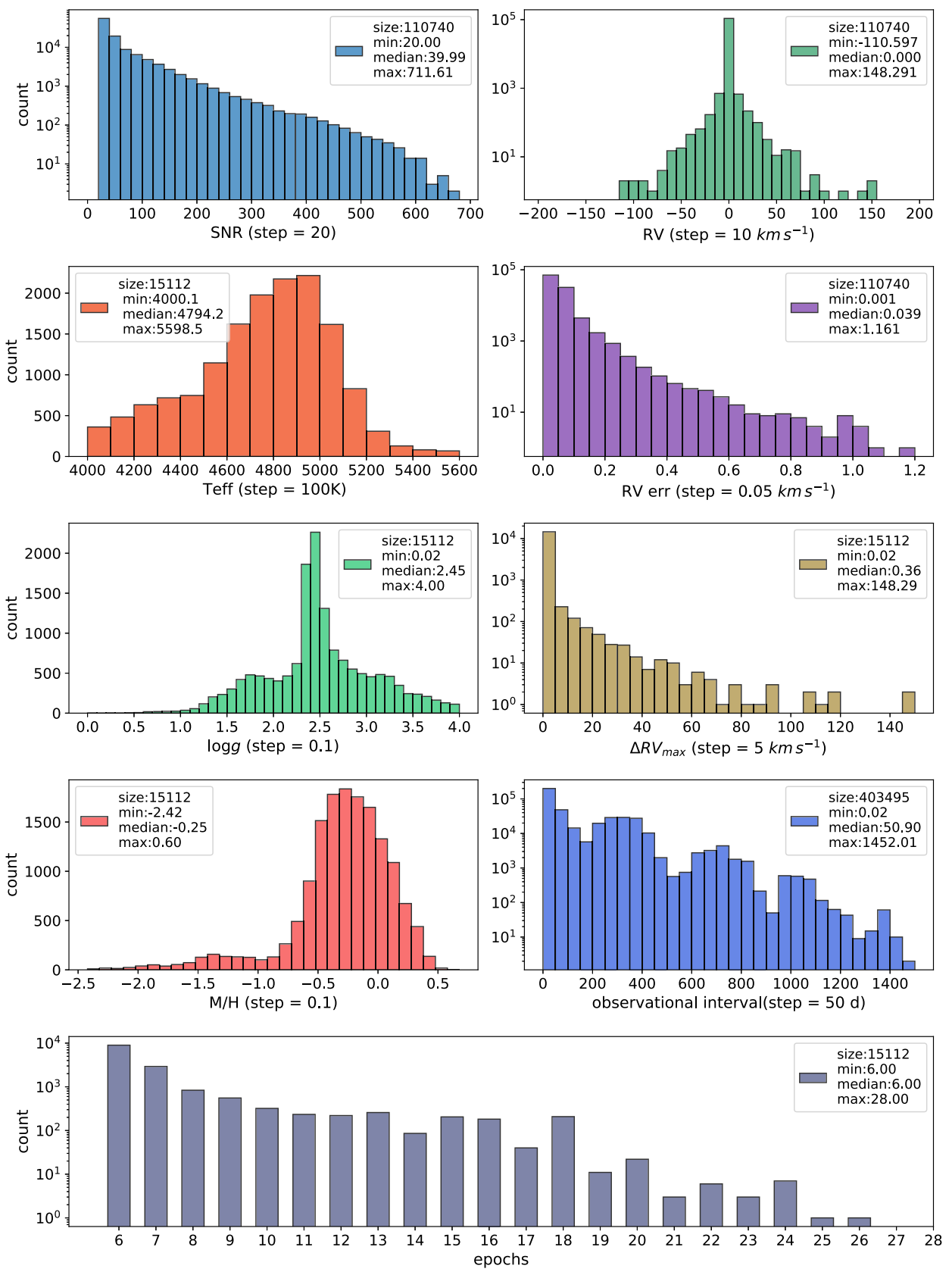

**Figure 16.** Basic information of the selected red giant samples.

### 5.2 Mean measurement error of RVs

A simple test was carried out to roughly estimate the spectroscopic binary fraction by using equation (8). Here $v_i$ and $\varepsilon_i$ are the RV and the RV error at epoch $i$ for a given observational source.

We varied $a$ from 1.0 to 6.0 in steps of 0.5 and recorded the corresponding binary fraction, as shown in Fig. 17. The directly detected $f_{\rm bin}$ shown in the figure exceeds 0.9 when $a = 3.0$, which seems unreasonably high. As summarized in (F. Luo et al. 2021), most main-sequence stars do not have a so high $f_{\rm bin}$, let alone red giant binaries. Therefore, the RV measurement error provided by the APOGEE DR16 catalogue may be underestimated. We need to reassess the RV measurement error of these samples.

It is assumed that the left part of the $\Delta RV_{\rm max}$ distribution curve from observational data was contributed by single stars. We generated single-star samples with different mean measurement errors of RV $\varepsilon$. The parts to the left of the observed peak of these samples were intercepted to compare with the corresponding observed curve utilizing the least square approach, and the minimum value emerged when $\varepsilon = 0.074$. Then we adopted $0.074\ {\rm km\,s^{-1}}$ as the uniform RV measurement error of all observational samples.

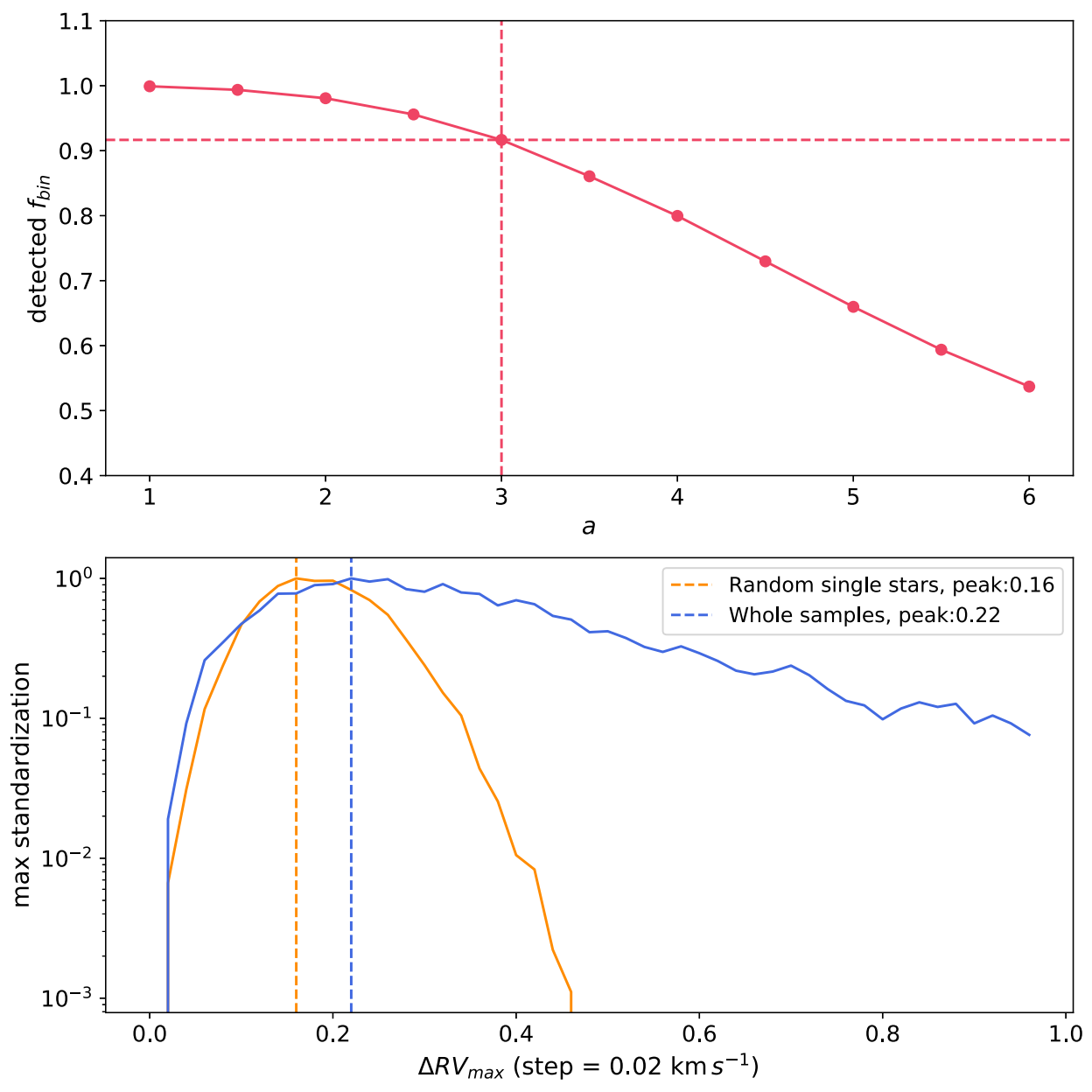

**Figure 17.** Estimating the error of RV through generations of single-star samples with different $\varepsilon$.

### 5.3 Lower limits of orbital period detection

Samples whose $2.2 < \log g < 2.8$ were not included since this area may contain many red clumps. We divided the remaining 7607

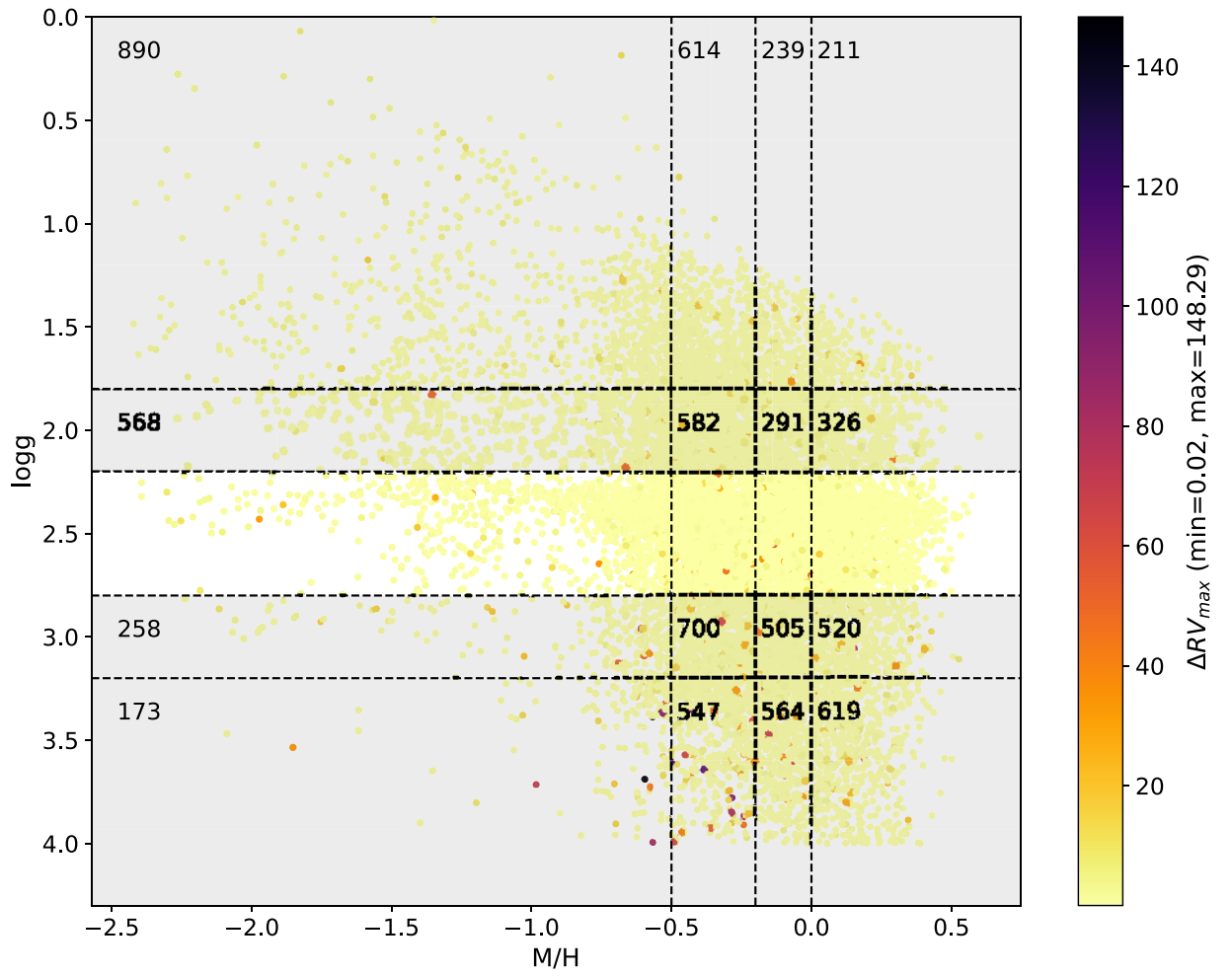

**Figure 18.** Meshes according M/H and logg, the number in each block is the number of observational sources included.

samples into 16 parts according to M/H and logg (see Fig. 18). We estimate minimum detectable orbital periods as a function of surface gravity by imposing that the primary radius not exceed its Roche lobe-limited separation. Adopting a representative primary mass $m_1$ and mass ratio $q$, a simple order-of-magnitude estimate is obtained:

$$P = \left(\frac{4\pi^2 R^3}{Gm_1\,(1+q)}\right)^{\frac{1}{2}}, \quad R = \sqrt{\frac{Gm_1}{10^{\mathrm{logg}}}}. \tag{6}$$

Estimates of the lower limits of $P$ for samples whose $0 < \mathrm{logg} < 1.8$ is 41.39 (d), $1.8 < \mathrm{logg} < 2.2$ is 7.07, and $2.8 < \mathrm{logg} < 4.3$ is 1.5. These limits are used to define the period ranges in subsequent Monte Carlo simulations.

### 5.4 Samples that mimic the real data

We obtained 1959 sources' primary masses by cross-matching with the star catalogue published by M. Ness et al. (2016). In the MCMC simulations we randomly draw $m_1$ evenly from these 1959 masses to generate the mock samples using equation (1). From Fig. 16 we can see the maximum of the observation interval between any pairs of observations is about 1452 d. It means that we can not detect the binaries with orbital periods longer than 1452 d from these RGB data. Then the upper limit of the detected range of $P$ was determined as 1400 d. Except $m_1$ and $P$, the detection ranges of other parameters were defined as the same as Table 2.

135 groups of mock samples were generated with different combinations of sample size, epochs, $\pi$, and $f_{\mathrm{bin}}$, as depicted in Fig. 20. The minimum and maximum orbital periods in these groups were 1.5 and 1400 d. The green, red, and blue lines in the figure represent the true value of $\pi = -1.5$, 0, and 1.5, respectively. Each cross-point of two dotted lines indicates the true value position of a group of mock data. Our analysis reveals that $f_{\mathrm{bin}}$ estimates demonstrate high accuracy across nearly all test cases. However, $\pi$ predictions exhibit reduced accuracy when $f_{\mathrm{bin}} = 0.1$, irrespective of sample size or number of observational epochs. As $f_{\mathrm{bin}}$ increases to 0.3 or higher, the results progressively converge towards their respective true values. Overall, $f_{\mathrm{bin}}$ estimations exhibit greater stability and accuracy compared to $\pi$ estimates. The precision of $\pi$ constraints improves with increasing data quantity and quality.

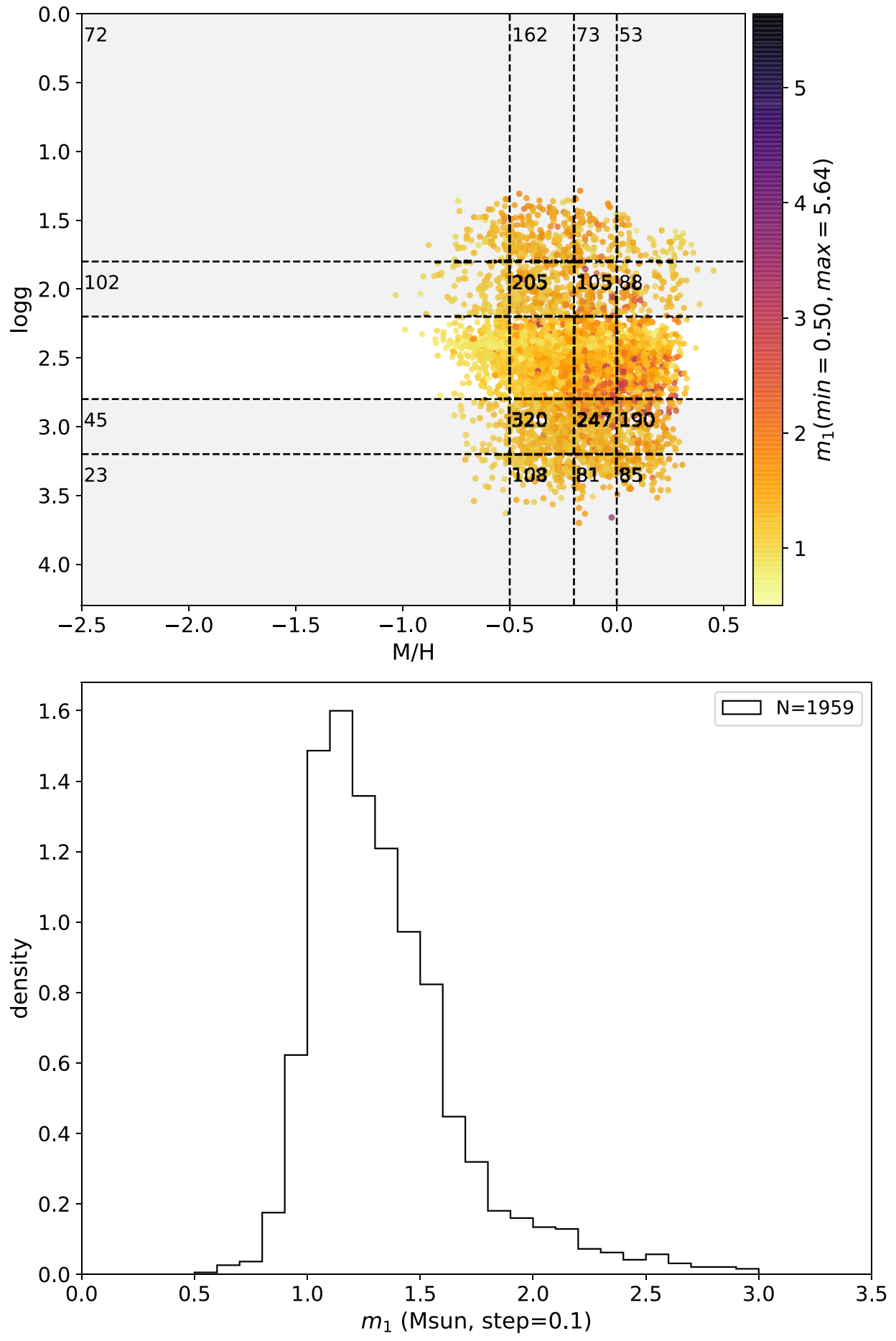

**Figure 19.** The obtained 1959 primary masses from M. Ness et al. (2016).

### 5.5 Results

We calculated all the 16 subsets of real data displayed as Fig. 18. The detection range of $\pi$ was $[-2, 2]$ with step 0.1, $f_{\mathrm{bin}}$ was [0,1] with step 0.01. The results of $\pi$ and $f_{\mathrm{bin}}$ with logg and M/H are displayed in Fig. 21. Points in each error bar are peak positions. The lower and upper limits of error bars represent the 15th and 85th percentiles. Lengths of error bars were defined as the half value of difference between the 15th and 85th percentiles, so some upper limits of $\pi$ were beyond the detection range.

From the figure, we can find that $\pi$ did not change significantly with logg and M/H; the error bars are also wide, especially when the corresponding $f_{\mathrm{bin}}$ is small. According to the above-mentioned simulations, we need more data to obtain precise predictions of $\pi$, but the $f_{\mathrm{bin}}$ results were trustworthy. The max $f_{\mathrm{bin}}$ is about 0.45 when $3.2 <= \mathrm{logg} < 4.3$ and $2.5 <= \mathrm{M/H} < -0.5$, and the min is about 0.05 when $1.8 <= \mathrm{logg} < 2.2$ and $0.0 <= \mathrm{M/H} < 0.6$. Generally, the $f_{\mathrm{bin}}$ estimations show a tendency of decreasing with the decrease of logg and decreasing with the increase of M/H.

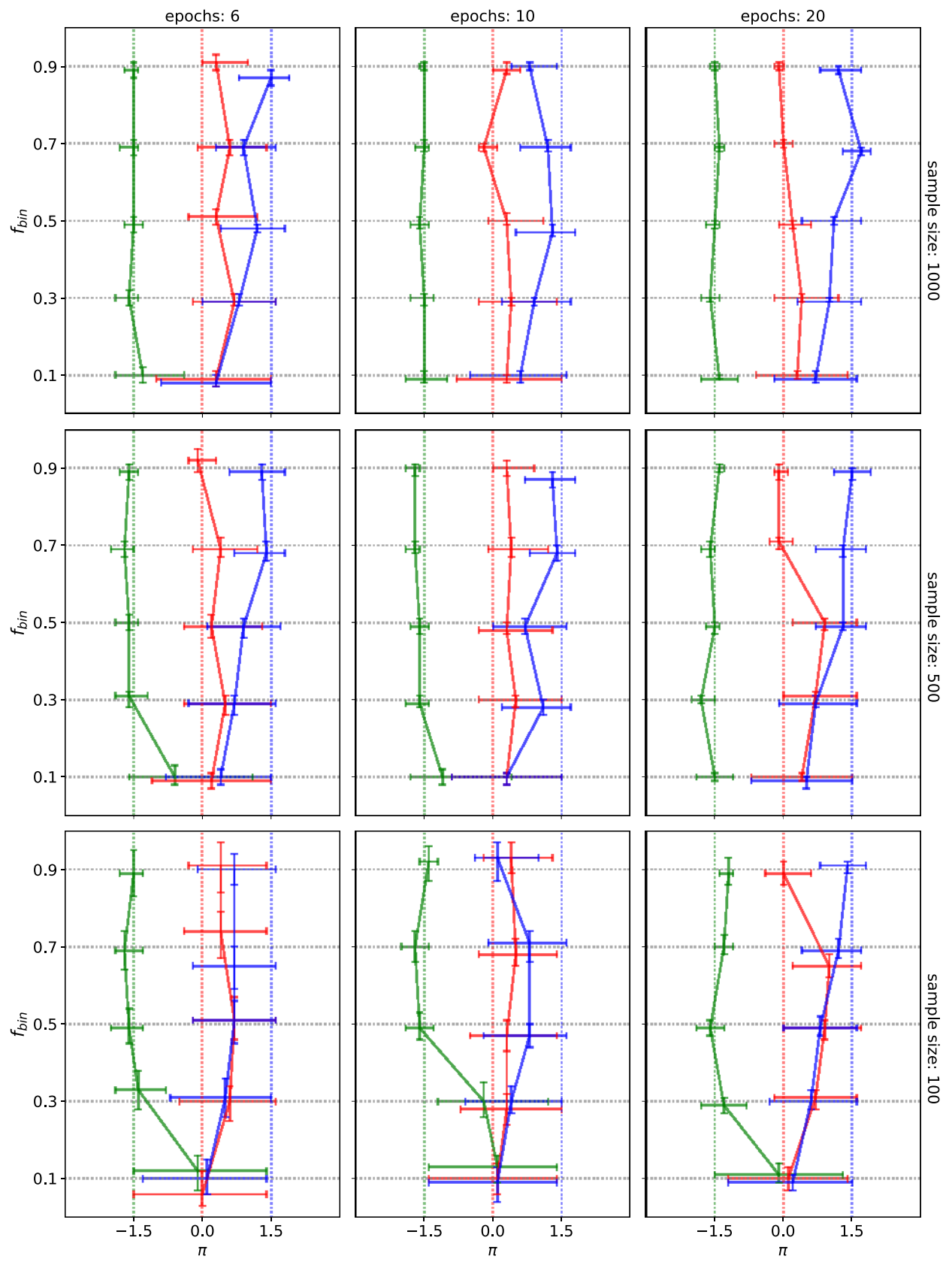

**Figure 20.** Simulated experiments with random samples that mimic the real data. Each cross-point of two dotted lines indicated the true value position of a group of mock data. Green, red, and blue lines represent the true $\pi$ of $-1.5$, 0, and 1.5, respectively.

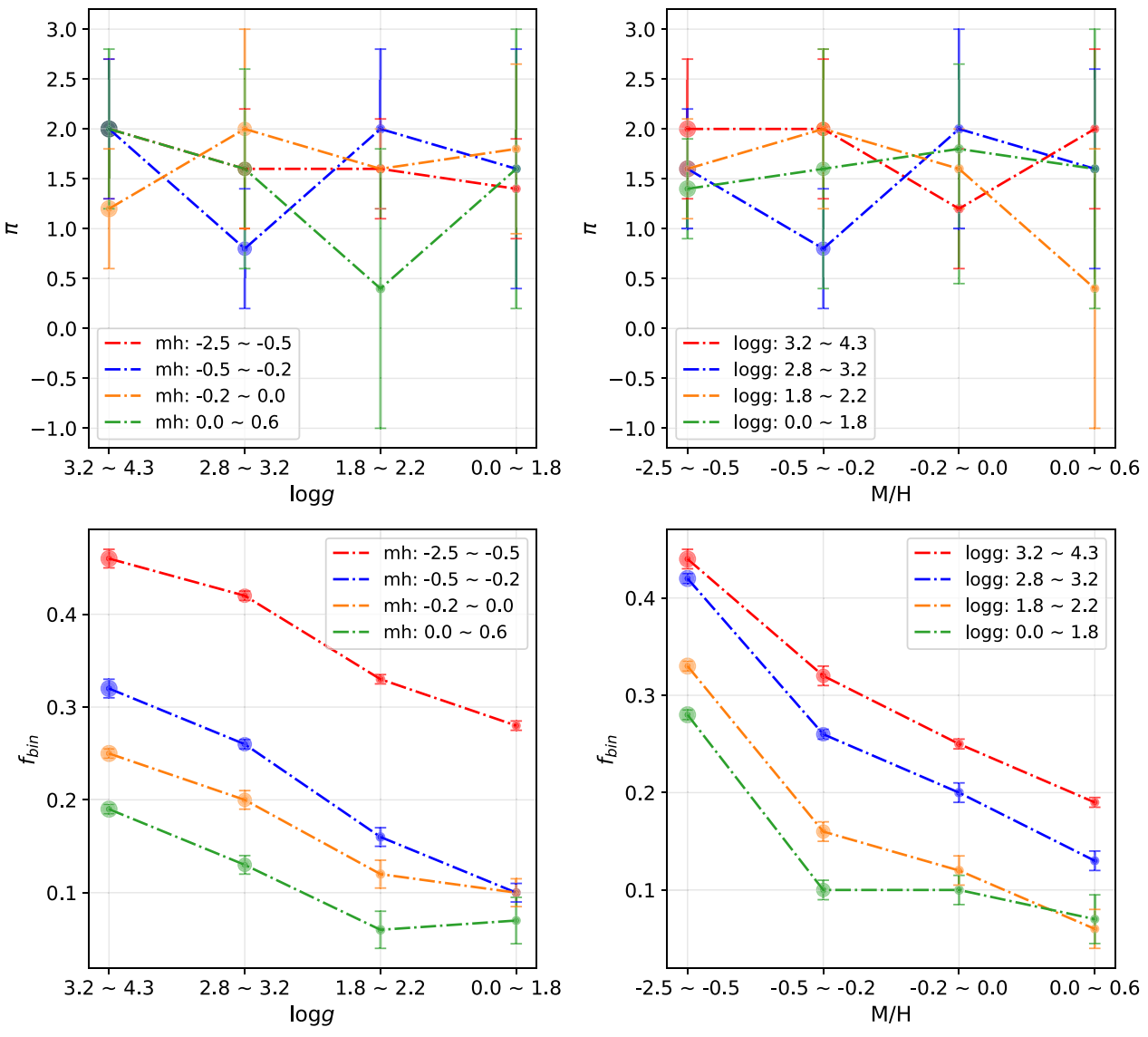

**Figure 21.** The $\pi$ and $f_{\rm bin}$ vary with logg and M/H. The size of the circle area at each point represents the sample size of that subset.

## 6 DISCUSSION AND CONCLUSION

### 6.1 The tendency of $f_{\rm bin}$

The $f_{\rm bin}$ results derived from APOGEE DR16 red giant samples by DVCD align with previous studies (C. Badenes et al. 2018; Y. Huang, H.-b. Yuan & Y. Huang 2020; C. N. Mazzola et al. 2020). The trend of $f_{\rm bin}$ decreasing with logg (Fig. 21) in red giants implies that some binary systems may be disrupted during evolutionary processes, or the primary may leave the RGB area in the HRD (Hertzsprung–Russell diagram), among other reasons.

For one thing, during the inflation of primary stars, the orbital angular momentum may be converted into the rotational angular momentum (R. E. Wilson & R. Stothers 1975; P. M. S. Namboodiri & R. K. Kochhar 1990; J. G. Greenhill, D. K. Galloway & J. R. Murray 1999; J. Zhang, S.-B. Qian & B. Soonthornthum 2009a; J. Zhang, S. B. Qian & Boonrucksar 2009b; J. F. Sepinsky, B. Willems & V. Kalogera 2011; D. Gerosa et al. 2015). The separation between the two components will become smaller, leading to a certain probability of Roche lobe overflow (A. R. King & C. Done 1993; D. R. Gies 2012; A. Vigna-Gómez et al. 2020) or CE (Common Envelope) events (E. Berzin et al. 2021; L. Jiang et al. 2021; R. Schreier et al. 2021). Finally, if the two stars merge into a single star, the binary system is devastated (N. Ivanova & P. Podsiadlowski 2002; K. Exter 2010; A. S. Jermyn & M. Cantiello 2020; D. Nakauchi, K. Inayoshi & K. Omukai 2020).

For another, if there is a non-negligible mass loss in a binary system (N. R. Evans et al. 2018; H. E. Brinkman et al. 2019; E. J. Farrell et al. 2020; D. N. Doikov & A. V. Yushchenko 2021; S. Ekström 2021; O. Pejcha et al. 2021; H. Sana 2022), whether due to stellar winds from the primary or secondary, the separation between the two components will increase along with the orbital period. The RV variation of such binary systems becomes less detectable until they can no longer be observed. Binary models based on radial velocities may regard these binaries as single stars, resulting in an apparent decrease in $f_{\rm bin}$.

The above-mentioned analysis of the $f_{\rm bin}$ tendency is from a statistical point of view. Although the tendency consisted of the above-mentioned theoretical analysis, the results require multi-party authentication. Evidence of how this trend arose requires further research. Particularly, the observations of ‘star by star’ are very helpful in demonstrating this phenomenon.

### 6.2 The DVCD method

We propose a new method, DVCD, for binary statistical research. The direct inspiration for this algorithm comes from two previous algorithms: the hierarchical Bayesian model (HBM; C. Liu et al. 2014; S. Gao et al. 2017; Z.-J. Tian et al. 2018) and the algorithm proposed by H. Sana et al. (2013, hereafter S13).

The main equation of HBM is

$$p(\Delta \mathrm{RV}) = f_{\mathrm{B}} p_{\mathrm{B}}(\Delta \mathrm{RV} \mid \sigma_{\mathrm{RV}}, \Delta \mathrm{t}, \mathrm{M_B}) + (1 - f_{\mathrm{B}})\, p_{\mathrm{S}}(\Delta \mathrm{RV} \mid \sigma_{\mathrm{RV}}), \tag{7}$$

where $f_B$ is the intrinsic binary fraction, $\sigma_{\rm RV}$ the measurement error of radial velocities. $\Delta t$ represents the time interval between two observational epochs. $p_{\rm B}(\Delta \mathrm{RV} \mid \sigma_{\rm RV}, \Delta \mathrm{t}, \mathrm{M_B})$ and $p_{\rm S}(\Delta \mathrm{RV} \mid \sigma_{\rm RV})$ denote probabilities of obtaining $\Delta$RV under the assumptions of binaries and single stars, respectively (Z.-J. Tian et al. 2018). $\mathrm{M_B}$ is a binary model using equation (1) to randomly draw simulated RVs containing all orbital parameters (Table 1) and possible hyper-parameters (Table 3).

In our simulation experiment, we found that the $f_{\rm bin}$ results obtained by the HBM algorithm were relatively accurate, but they often failed to provide the distribution information of the orbital parameters. After analysis, the main reason for this phenomenon lies in the criteria $\sigma_{\rm RV}$ or $\Delta {\rm RV}_{\rm max}$ used by HBM, both of which are single statistical quantities. They can reflect the magnitude of the RV variations, but they lose the information about the observation time and the trend of how RV changes. If the number of random samples during the algorithm execution process is increased, the error range of parameters such as $\pi$ and $\kappa$ can be reduced (though still not accurate enough). But the running time of HBM will significantly increase, making it difficult to complete large-scale statistical analysis.

The main equations of the S13 method are

$$\frac{|v_i - v_j|}{\sqrt{\varepsilon_i^2 + \varepsilon_j^2}} > a \quad \text{and} \quad |v_i - v_j| > C, \tag{8}$$

$$\Xi' = P_{\rm KS}(\Delta {\rm RV}) \times P_{\rm KS}(\Delta {\rm HJD}) \times B\left(N_{\rm bin}, N, f_{\rm bin}^{\rm simul}\right). \tag{9}$$

$a = 4.0$ was adopted in S13 and a global merit function ($\Xi'$) is constructed by multiplying the two KS probabilities obtained for the $\Delta$RV and $\Delta$HJD distributions and the Binomial Probability. Where $\Delta$RV are those satisfying equation (8), and $\Delta$HJD are the corresponding observational intervals. $f_{\rm bin}^{\rm simul}$ is the $f_{\rm bin}$ detected in simulated samples; $N_{\rm bin}$ is the number of binaries detected in observational data by equation (8); $N$ is the sample size.

A hidden error comes from the third term in equation (9), the Binomial probability $B\left(N_{\rm bin}, N, f_{\rm bin}^{\rm simul}\right)$. The accuracy of this term depends on the estimation of $N_{\rm bin}$. However, the $N_{\rm bin}$ is determined by equation (8), which is sensitive to the parameters $a$ and $C$. Different samples (such as spectral types) and observational conditions (such as measurement errors) may require different $a$ and $C$. The choice of $a$ and $C$ is somewhat subjective, which will lead to different $N_{\rm bin}$ and then affect the final results.

Another potential mistake is the issue of the degree of variation in RVs. In the S13 method, as long as there is any pair of RVs from a certain source that conforms to formula (8), it will be marked as a binary star. Among all the pairs of RVs satisfying formula (8), the one that is most likely to be selected is $\Delta {\rm RV}_{\rm max}$, which is the difference between the maximum and minimum values in the RVs. Due to the existence of measurement error, the $\Delta {\rm RV}_{\rm max}$ from a single star will behave more and more like a binary as the number of observational epochs increases.

The DVCD algorithm effectively addresses these limitations. First, we do not need to introduce subjective quantities like $a$ and $C$ to preprocess the original data. Second, the distribution of the adopted criterion $\Delta$AJRV for single stars is always $N(0, \sqrt{2\varepsilon})$ and does not change with the number of observations. The accuracy of the calculated results will increase as the number of observations increases. When there is insufficient original data for us to obtain accurate predictions of orbital period distributions, the $f_{\rm bin}$ results are still accurate (Fig. 20).

Meanwhile, we conducted a test on the running times of these three algorithms. For the sake of caution and accuracy, during the execution of the DVCD algorithm, we did NOT use a cache as described before, all random sampling was calculated in real time. The number of random sampling times was set to $10^4$ for all of them. It can be found from Fig. 22 that both S13 and DVCD can produce effective constraints on $f_{\rm bin}$, but the HBM indicates a systematic underestimation. Our experimental results also demonstrate that DVCD achieves a $10^4$ to $10^5$-fold improvement in computational efficiency compared to the S13 method, and $10^2$ to $10^3$ to HBM. This gives DVCD a unique advantage when dealing with a large number of samples. We can quickly explore the different distributions or ranges of the track parameters. If caching is used, the calculation will be faster.

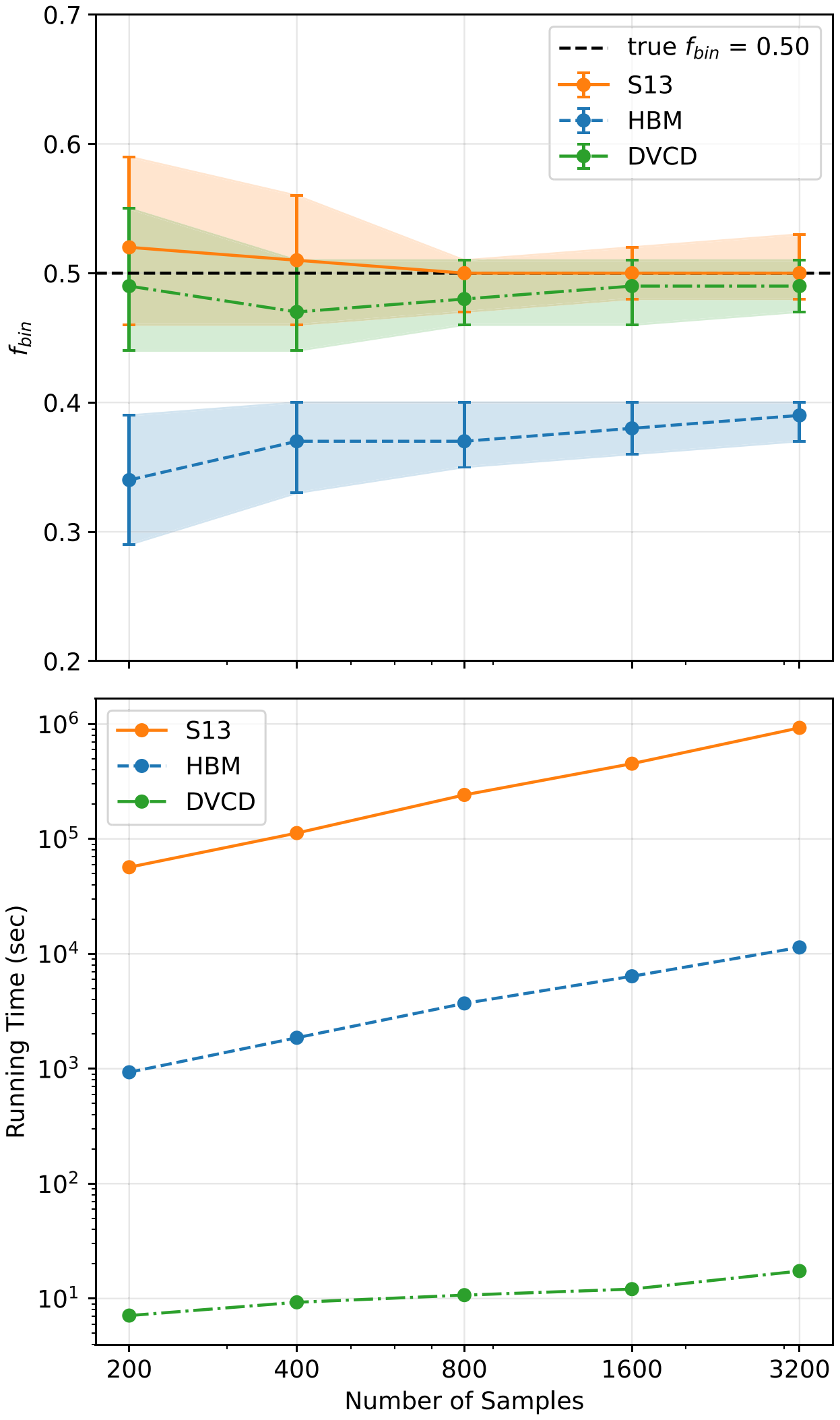

**Figure 22.** Comparisons of S13, HBM, and DVCD. $(\pi, \kappa, \eta, \varepsilon, f_{\rm bin}) = (-1.5, 0, 0, 1.0, 0.5)$, epochs = 6, and $10^4$ random samplings in the simulation procedure.

The DVCD algorithm exhibits reliable accuracy and efficiency, making it a promising tool for large-scale statistical analysis of binary-star systems. However, the DVCD algorithm was developed based on the previous algorithms by breaking through some possible limitations. It is directly rooted in the explorations of its predecessors.

## ACKNOWLEDGEMENTS

This work was supported by National Astronomical Observatories, Chinese Academy of Sciences (No. E4TG2001).

We thank the APOGEE team for making the DR16 data publicly available, and we sincerely appreciate the guidance and constructive comments on scientific applications of the DVCD algorithm from Prof. Zhao Yongheng and Prof. Liu Chao. We also acknowledge helpful discussions with colleagues at the National Astronomical Observatory of China, NAOC.

## DATA AVAILABILITY

The data underlying this article are available in the article and in its online supplementary material.

This paper has been typeset from a TEX/LATEX file prepared by the author.